\documentclass[%
superscriptaddress,
preprintnumbers,
nofootinbib,
 amsmath,amssymb,
 aps,
prd,
twocolumn,
longbibliography
]{revtex4-1}
\usepackage[normalem]{ulem}
\usepackage{hyperref}
\usepackage[dvipsnames]{xcolor}
\usepackage{comment}
\usepackage{siunitx}
\usepackage{graphicx}
\usepackage{dcolumn}
\usepackage{bm}
\newcommand{\qhat}{\hat q}

\newcommand{\dd}[2][]{\mathrm d^{#1}{#2}\,} 
\newcommand{\vb}[1]{\mathbf{#1}}
\newcommand{\Teps}{T_\varepsilon}
\newcommand{\tauT}{\tau_{\mathrm{BMSS}}}
\newcommand{\tauR}{\tau_R}

\newcommand{\pdv}[2][]{\frac{\partial{#1}}{\partial{#2}}}
\newcommand{\Conetwo}{\mathcal {C}^{1\leftrightarrow 2}}
\newcommand{\Ctwotwo}{\mathcal{ C}^{2\leftrightarrow 2}}

\newcommand{\CR}{C_{\mathrm{R}}}
\newcommand{\CA}{C_{\mathrm{A}}}
\newcommand{\CF}{C_{\mathrm{F}}}
\newcommand{\NC}{N_{\mathrm{C}}}

\newcommand{\tform}{t^{\mathrm{form}}}
\newcommand{\Gammael}{\Gamma^{\mathrm{el}}}
\newcommand{\omegamin}{\omega_1}

\usepackage{siunitx}

\begin{document}
\title{Jet energy loss in anisotropic plasmas meets limiting attractors}

\preprint{MIT-CTP/6013}

\author{Kirill Boguslavski}
\email{kirill.boguslavski@subatech.in2p3.fr}
 \affiliation{SUBATECH UMR 6457 (IMT Atlantique, Université de Nantes, IN2P3/CNRS), 4 rue Alfred Kastler, 44307 Nantes, France}
 \affiliation{Institute for Theoretical Physics, TU Wien, Wiedner Hauptstrasse 8-10, 1040 Vienna,
Austria}

\author{Lucas Hörl}
\email{lhoerl@ethz.ch}
 \affiliation{Department of Physics, ETH Zürich, 8093 Zürich, Switzerland}
 \affiliation{Institute for Theoretical Physics, TU Wien, Wiedner Hauptstrasse 8-10, 1040 Vienna,
Austria} 
\author{Florian Lindenbauer}
\email{flindenb@mit.edu}
\affiliation{MIT Center for Theoretical Physics -- a Leinweber Institute, Massachusetts Institute of Technology, Cambridge, MA 02139, USA}
\affiliation{Institute for Theoretical Physics, TU Wien, Wiedner Hauptstrasse 8-10, 1040 Vienna,
Austria}

\date{\today}

\begin{abstract}
We consider the energy loss of a high-energy parton in the early anisotropic plasma in heavy-ion collisions. Working in the harmonic approximation, we compute the change in the mean energy of an emitted gluon in the presence of an anisotropic background, characterized by anisotropic jet quenching parameters $\hat q_{x}\neq \hat q_{y}$. Evaluating the resulting integrals numerically, we compare with isotropic media, and obtain a simple pocket formula to estimate the impact of anisotropy on the mean emitted gluon energy, which is generally small. We then combine our results with the values of the jet quenching parameter extracted from QCD kinetic theory simulations and show that the medium length dependence of this mean energy loss exhibits the characteristics of limiting attractors, which can be obtained by extrapolating to zero and infinite coupling. Our study thus relates energy loss of jet partons to universal dynamics of anisotropic plasmas.

\end{abstract}

\maketitle

\section{\label{sec:level1}Introduction}

Relativistic heavy-ion collision experiments produce an exotic state of deconfined matter that consists of quarks and gluons, referred to as the quark-gluon plasma. By studying its properties in these collision experiments, we aim to learn more about the strong interaction (QCD) under extreme conditions, both in and out of equilibrium
\cite{Busza:2018rrf}. These collisions create a QCD plasma in an initially far-from-equilibrium state, which quickly isotropizes, and a hydrodynamic description becomes applicable \cite{Berges:2020fwq}. 
Its properties can be probed by measuring the modification of jets
from which features of this exotic phase of QCD matter can be deduced.

While most studies focus on the hydrodynamical description of the plasma, the initial nonequilibrium stages encode interesting physics and may offer a unique laboratory for studying features of nonequilibrium QCD \cite{Berges:2020fwq, Schlichting:2019abc}. 
The plasma in this regime is neither sufficiently strongly nor weakly coupled, such that in practice, extrapolations of strongly-coupled
or weakly-coupled methods need to be performed (see, e.g., \cite{Heller:2012km, Keegan:2015avk, Berges:2020fwq}).
In this work, we focus on the weakly-coupled picture. There, the initial stages after the collision are dominated by classical gluon fields \cite{McLerran:1993ni, Gelis:2012ri}
until an effective kinetic description in terms of gluons and quarks becomes applicable \cite{Berges:2013eia, Baier:2000sb}. This QCD kinetic theory \cite{Arnold:2002zm} can be solved numerically to follow the system from an initial far-from-equilibrium state to isotropy, when a hydrodynamic description of the plasma becomes applicable \cite{Kurkela:2014tea, Kurkela:2015qoa, kurkela_2023_10409474, Gale:2013da, Heinz:2024jwu}.

Possible experimental probes of the pre-hydrodynamic dynamics consist of jets,
which are particle showers originating from highly energetic partons 
created in the initial collision.
While an energetic parton traverses the plasma, it interacts with the medium and loses energy, a phenomenon called \emph{jet quenching} \cite{Qin:2015srf, Apolinario:2022vzg, Mehtar-Tani:2025rty}. This energy loss is dominated by inelastic gluon emissions, for which several formalisms have been put forward to describe this process. For highly energetic partons, and in the often-employed harmonic approximation, the effect of the medium is encoded in a single jet quenching parameter $\qhat$, which has the physical interpretation of quantifying the momentum broadening of partons traversing the plasma.

Phenomenological studies of jets mostly focus on the hydrodynamic stage of the QGP evolution, ignoring or drastically simplifying the initial nonequilibrium stages in the plasma time evolution. However, recent works have indicated the possibility of sensitivity of jets to the initial stages \cite{Andres:2019eus, Adhya:2024nwx, Barata:2025agq, Barata:2025zku}.
Additionally, the  
medium parameter
$\hat q$, which characterizes jet momentum broadening and plays a crucial role in energy loss calculations \cite{JET:2013cls, Andres:2016iys, Andres:2019eus, Mehtar-Tani:2021fud, JETSCAPE:2021ehl, Xie:2022ght, JETSCAPE:2024cqe, Pablos:2025cli}, has further been extracted in the early Glasma and kinetic theory stages \cite{Ipp:2020mjc, Ipp:2020nfu, Carrington:2020sww, Carrington:2021dvw, Carrington:2022bnv, Avramescu:2023qvv, Boguslavski:2023waw, Boguslavski:2024ezg, Boguslavski:2024kbd}.
In particular, it has been reported to be anisotropic and large compared to the later hydrodynamic values. We will discuss how this parameter enters the calculation of jet energy loss in Section~\ref{sec:1}.

As the community keeps moving forward towards the initial stages, partly triggered by the new smaller collision systems studied at LHC, several calculations have been performed to include anisotropies in this formalism \cite{Hauksson:2021okc, Hauksson:2023dwh, Hauksson:2023, Lindenbauer:2025ctw, Barata:2024bqp, Silva:2025dan, Barata:2025zku}.
In this paper, we consider the effect of the plasma anisotropy on the energy spectrum, which is less differential but provides a simple and straightforward way to study the effect of the anisotropy.
In particular, we demonstrate that the moment of this energy spectrum, corresponding to the mean energy of one emitted gluon, shows features of the previously introduced limiting attractors \cite{Boguslavski:2023jvg}. We also study the effect of the anisotropy on the quenching weights, which are a way of quantifying the difference between the parton cross section in vacuum and in the presence of a medium \cite{Salgado:2003gb, Baier:2001whe}.

This paper is structured as follows. In Section~\ref{sec:gluonemission}, we discuss how the probability of a single gluon emission is calculated in the harmonic approximation. We generalize the isotropic calculation to an anisotropic plasma, and provide numerical results for the mean energy of an emitted gluon in an anisotropic static plasma. We further study its impact on the quenching weights. In Sec.~\ref{sec:kinetictheory}, we then use the previous results to compute the mean gluon energy of jets traversing the nonequilibrium plasma described with QCD kinetic theory. We also discuss the relation to limiting attractors. We conclude in Section \ref{sec:conclusion}. The appendices provide explicit details on the DGLAP splitting functions used in the main text, and on the numerical techniques employed in this paper, in particular on how we extrapolate the integration boundaries.

\section{Gluon emission in the harmonic approximation\label{sec:gluonemission}}
\label{sec:1}

\subsection{Isotropic medium}
The energy loss of energetic partons is dominated by inelastic gluon emissions.

These are triggered by elastic kicks with the medium, which result in a change of the transverse momentum of the parton, characterized by the jet quenching parameter $\qhat=\frac{\dd\langle p_\perp^2\rangle}{\dd t}$. The typical momentum of these transverse kicks is given by the screening mass $m_D$. The formation time $\tform$ of such an emission process increases with the energy $\omega$ of the emitted gluon, $\tform\sim\sqrt{\omega/\qhat}$ \cite{Schlichting:2019abc}. While these small momentum kicks off the medium occur at a rate of $\Gammael\sim\qhat/m_D^2$, for the emission of a sufficiently energetic gluon, the formation time becomes longer than the time between two elastic scatterings. This happens when $\omega \gtrsim m_D^4/\qhat$, and leads to a quantum-mechanical interference effect known as the Landau-Pomeranchuk-Migdal (LPM) suppression \cite{Landau:1953ivy, Migdal:1956tc}, first developed for QED and later for QCD \cite{Zakharov:1996fv, Zakharov:2000, BAIER1997291, Baier:1996sk}.
In this regime, the gluon emission spectrum $\dd I/\dd\omega$, which represents the probability of the energetic parton to emit a gluon with energy $\omega$ during its time evolution,
is given by
\cite{Zakharov:2000}
\begin{equation}
    \omega\mathcal{P}(\omega)=\omega \frac{\mathrm d}{\dd\omega} (I - I_{\text{vac}}) = \frac{\alpha_s x P_{\text{s} \rightarrow \text{g}}(x)}{\left[x(1 - x)\mathcal{E} \right]^2}\; \text{Re}\,\Lambda,
    \label{emissionrategeneral}
\end{equation}
with
\begin{multline}
    \Lambda=\int_{0}^{\infty} \dd {t'} \int_{t'}^{\infty} \dd t \ \nabla_{\mathbf{r}} \nabla_{\mathbf{r'}}[\mathcal{K}(\mathbf{r'}, t'; \mathbf{r}, t) \\- \mathcal{K}_{\text{vac}}(\mathbf{r'}, t'; \mathbf{r}, t)] \bigg|_{\mathbf{r}=\mathbf{r'}=0},
    \label{integralemissionrate}
\end{multline}
where $\alpha_s=\frac{\lambda}{4\pi N_\mathrm{C}}$
represents the strong coupling constant while $\lambda$ denotes the 't Hooft coupling. Throughout our numerical analysis, we use $N_\mathrm{C}=3$ for QCD.

The energy fraction of the emitted gluon compared to the total energy of the energetic parton $\mathcal E$ is denoted by $x={\omega}/{\mathcal{E}}$. The DGLAP splitting function $P_{s\rightarrow g}(x)$ depends
on the species $s$ of the initial particle, which can be either a gluon (g) or a quark (q), and is provided explicitly in Appendix~\ref{sec:dglap}. 
In Eq.~\eqref{integralemissionrate}, the functions $\mathcal{K}(\mathbf{r'}, t'; \mathbf{r}, t)$ and $\mathcal{K}_{\text{vac}}(\mathbf{r'}, t'; \mathbf{r}, t)$ are Green functions of a time-dependent Schrödinger equation in two dimensions,
\begin{equation}
    i \frac{\partial\Psi(\mathbf{r},t)}{\partial t} ={\mathcal{H}} \Psi(\mathbf{r},t),
\end{equation}
satisfying the initial condition 
\begin{align}
    \mathcal{K}(\mathbf{r_2}, t; \mathbf{r_1}, t) = \delta(\mathbf{r_2}-\mathbf{r_1}).
\end{align}
Here, $\mathcal{H}$ is the associated Hamiltonian, which can be decomposed into $\mathcal{H}=\delta E-i\Gamma_3$, where 
\begin{align}\label{eq:energy-difference}
    \delta E=\frac{p_r^2}{2x(1-x)\mathcal E}+\frac{m_s^2(t)}{2(1-x)\mathcal E}+\frac{m_g^2(t)}{2x\mathcal E}-\frac{m_s^2(t)}{2\mathcal E}
\end{align}
represents the energy difference between the gluon and the parton after the emission and the parton before the process, and $i\Gamma_3$ is a complex-valued potential. 
We have defined the effective momentum $\vb p_r = (p\vb k_\perp - k\vb p_\perp)/\mathcal E$.
This separation of the Hamiltonian is reminiscent of the usual split in kinetic and potential energy, where the latter is due to the interactions. 

The vacuum Green function $\mathcal{K}_{\text{vac}}(\mathbf{r'}, t'; \mathbf{r}, t)$ corresponds to the special case of vanishing interactions $\Gamma_3=0$, while $\mathcal{K}(\mathbf{r'}, t'; \mathbf{r}, t)$ includes the medium-dependent interaction potential $\Gamma_3$.
It is convenient to express $\Gamma_3$ in terms of the \emph{dipole cross section} or \emph{two-body potential} $\Gamma_2$ \cite{Zakharov:2000iz, Nikolaev:1993th}
\begin{multline}
    \Gamma_3(\mathbf{r},t)=\frac{1}{2}\Gamma_2(\mathbf{r},t) +\left(\frac{C_s}{\CA}-\frac{1}{2}\right) \Gamma_2(x\mathbf{r},t)\\+\frac{1}{2}\Gamma_2\left((1-x)\mathbf{r},t\right),
\end{multline}
where
\begin{align}
    \Gamma_2(\mathbf{r},t) = \int \frac{\dd[2]{\vb q_\perp}}{(2\pi)^2}\mathcal{C}( \vb q_\perp, t) \left(1-e^{i\mathbf{q}_\perp\cdot \mathbf{r}}\right)
    \label{helper-potential}
\end{align}
denotes the dipole cross section, and $\mathcal{C}\left(\vb q_\perp\right)=(2\pi)^2 \frac{\dd \Gamma}{\dd[2]{\vb q_\perp}} $ is the probability of receiving a transverse momentum kick $\vb q_\perp$. The Casimir factors are given by $\CA = \NC$, $\CF=(\NC^2-1)/(2\NC)$. At leading logarithmic approximation, and for an isotropic system, the full Hamiltonian reduces to that of a harmonic oscillator,
\begin{align}
    \hat{\mathcal{H}} \approx \frac{p^2_{\mathbf{r}}}{2M} + \frac{1}{2}M \Omega^2(t)r^2,
    \label{iso_harmonic_osc_approx}
\end{align}
with the complex and time-dependent frequency \cite{BAIER1997291} given by
\begin{align}
    \Omega^2(t) = -i \frac{(1-x) + x^2\frac{C_{\text{s}}}{\CA}}{2x(1-x) \mathcal E} \hat{q}(t) \overset{x\ll 1}{\approx} - \frac{i}{2\omega} \hat{q}(t).
    \label{eq:omega_squared_relation_qhat}
\end{align}

The jet quenching parameter $\hat q$ entering here encodes the transverse momentum broadening of a gluon.
Here, we have employed the additional approximation $x\ll 1$, corresponding to the emitted gluon having a small energy $\omega$ compared to the jet energy $\mathcal E$. Within this approximation, $M=x(1-x)\mathcal{E} \approx \omega$, and the dependence on the jet energy $\mathcal E$ drops out. Additionally, we have assumed $\omega \gg m_s$, i.e., that the energy of the emitted gluon is much larger than the screening scales of the plasma, such that we may neglect the mass-dependent corrections appearing in the energy difference \eqref{eq:energy-difference}.
Since we work in the LPM regime, where the emitted gluon energy is sufficiently large, $\omega\gg m_D^4/\qhat$, this is well justified. For instance, for a thermal system where $m_D^4/\qhat\sim T$, this implies that the emitted gluon is energetically well separated from the medium, $\omega \gg T$.
To summarize, we thus work in the limit $\max(m_s,m_s^4/\qhat)\ll \omega \ll \mathcal E$.

The (isotropic) jet quenching parameter $\qhat$ appearing in the frequency \eqref{eq:omega_squared_relation_qhat} is related to the probability $\mathcal C(q_\perp)$ via
\begin{equation}
    \hat{q}(t) = \int^{\Lambda_\perp} \frac{\dd[2]{\vb q_{\perp}}}{(2\pi)^2}\, \mathcal{C}\left(q_\perp,t\right) \left(q_{\perp}\right)^2
    \label{jet-quenching-parameter-iso}
\end{equation}
and emerges in the expansion of Eq.~\eqref{helper-potential} for small values of $\mathbf{q}_\perp\cdot \mathbf{r}$. In practice, Eq.~\eqref{jet-quenching-parameter-iso} needs to be supplemented by a momentum cutoff $\Lambda_\perp$ to render $\hat q$ finite. However, a closer inspection of the small-distance behavior of the potential \eqref{helper-potential} reveals that a cutoff is not necessary if, in addition to the quadratic behavior, a logarithmic correction related to $\hat q$ is taken into account \cite{Altenburger:2025iqa}. However, the simpler harmonic approximation \eqref{iso_harmonic_osc_approx} is often used, which only assumes a quadratic dependence on $r$. For simplicity, we will follow here the same strategy and employ a specific cutoff for $\qhat$, which we will detail later. We note that in the limit $x\ll 1$, the splitting function appearing in Eq.~\eqref{emissionrategeneral} can also be simplified to
\begin{align}
     x P_{\text{s} \rightarrow \text{g}}(x) \approx 2 \CR,
\end{align}
where $\CR=\CA$ for an initial gluon, and $\CR=\CF$ for a quark. Using the harmonic oscillator approximation \eqref{iso_harmonic_osc_approx}, we may then simplify the gluon emission spectrum \eqref{emissionrategeneral} using the Green's function of the time-dependent one-dimensional harmonic oscillator \cite{doi:10.1142/7305},
\begin{align}
    &\mathcal K^{1D}(x',t'; x,t)=\sqrt{\frac{M}{2\pi i \,S(t',t)}}\\
    &\quad\times \exp\left[\frac{iM}{2S(t',t)}\left(x'{}^2\dot{S}(t',t)-x^2\dot S(t,t')-2xx'\right)\right]\nonumber
\end{align}
to \cite{Arnold:2008vd}
\begin{multline}
    \omega \mathcal{P}_\text{iso}(\omega) = -\frac{2\alpha_s \CR}{\pi} \mathrm{Re} \int_{0}^{\infty} \dd{t'} \int_{t'}^{\infty} \dd t  \\\left[\frac{1}{S^{2}(t,t')} - \frac{1}{(t-t')^2} \right],
        \label{iso_integral}
\end{multline}
where $S(t,t')$ is a solution to the following differential equation with boundary conditions
\begin{equation}
    \Ddot{S} = - \Omega ^2 (t) S; \quad S(t=t',t')=0, \quad \Dot{S}(t=t',t')=1.
    \label{differential_equation_with_boundary}
\end{equation}
The crucial insight from Ref.~\cite{Arnold:2008vd} is that the $1/S^2$-term appearing in Eq.~\eqref{iso_integral} can be written as a total derivative, reducing the inner integral to the boundary terms,\footnote{These functions are conventionally called $S$ and $C$ because they reduce to sine and cosine functions for a static brick in the interior, see Eq.~\eqref{eq:S_a_functions}.
}
\begin{align}
    \frac{1}{S^2}=-\partial_t\left(\frac{C}{S}\right), \qquad C(t,t')=-\partial_{t'}S(t,t').
    \label{simple-formula-trick}
\end{align}
This allows explicitly evaluating Eq.~\eqref{iso_integral} for any time-profile of $\hat q(t)$, to obtain 
\begin{subequations}
\begin{align}
   \omega \mathcal{P}_\text{iso}(\omega) = \frac{2\alpha_s \CR}{\pi} \ln|c(0)|,
    \label{emission_rate_reformulated}
\end{align}
where the function $c(t)$ solves the differential equation
\begin{align}
    \Ddot{c}(t) = - \Omega^2(t) c(t); \quad c(t \rightarrow \infty) = 1, \quad \dot{c}(t \rightarrow \infty) = 0.
    \label{simple_formula_differential_eq}
\end{align}
\end{subequations}
This is the \emph{simple formula} obtained in Ref.~\cite{Arnold:2008vd}. 
We would like to emphasize that the trick \eqref{simple-formula-trick} for the integrand of the gluon emission spectrum requires an isotropic jet quenching parameter to obtain the simple expression \eqref{emission_rate_reformulated}. In an anisotropic system, the integrand cannot be easily written as a total derivative, and hence, the appearance of an anisotropic (direction-dependent) jet quenching parameter $\qhat_{ij}$ requires an extension of the formalism, which will be the focus of the following sections.

\subsection{Generalization to an anisotropic plasma}\label{sec:generalization-anisotropic}
In order to take the plasma anisotropy into account, we promote the jet quenching parameter to a tensor, a common technique for anisotropic broadening \cite{Hauksson:2023, Boguslavski:2023waw, Barata:2024bqp, Barata:2025uxp, Barata:2025wnp, Barata:2025zku, Danhoni:2026gve},
\begin{align}
    \hat{q}_{ij}(t) = \int^{\Lambda_\perp} \frac{\dd[2]{\vb q_{\perp}}}{(2\pi)^2}\, \mathcal{C}\left(t,\vb q_\perp\right) q_{\perp}^i q_{\perp}^j.
    \label{jet-quenching-parameter-matrix}
\end{align}
From its definition, the jet quenching parameter $\qhat_{ij}$ is symmetric, and we work here in a basis in which it is diagonal.
Thus, we describe the anisotropy only with its two diagonal elements
\eqref{jet-quenching-parameter-matrix} $\hat{q}_x = \hat{q}_{xx}$ and $\hat{q}_y = \hat{q}_{yy}$ ,
representing the rates of change of the transverse momentum in the $x$ and $y$ direction, respectively, with the parton moving in the $z$-direction.

We will frequently compare to an isotropic system and study the respective deviations using the isotropized and relative jet quenching parameters
\begin{align}
    \hat q = \hat q_x + \hat q_y, && \Delta \hat q = \hat q_x -\hat q_y.\label{eq:qhat-comparison-iso}
\end{align}
In the anisotropic case, the Hamiltonian of the Schrödinger equation~\eqref{iso_harmonic_osc_approx} is given by
\begin{align}
    \hat{\mathcal{H}} = \frac{p_x ^2 + p_y ^2}{2M} + \frac{M}{2} \Omega_x ^2 (t) x^2 + \frac{M}{2} \Omega_y ^2 (t) y^2.
\end{align}
Here, we have generalized $\Omega_x^2=-i \frac{\hat{q}_x}{\omega}$,  $\Omega_y^2=-i \frac{\hat{q}_y}{\omega}$. For a two-dimensional anisotropic harmonic oscillator, the gluon emission spectrum~\eqref{iso_integral} becomes
\begin{multline}
    \omega \mathcal{P}_\text{aniso}(\omega)=-\frac{\alpha_s \CR}{\pi}\; \mathrm{Re} \int_{0}^{\infty} \dd{t'} \int_{t'}^{\infty} \dd t \\ \ \left[ \frac{1}{S_x^{1/2}S_y^{1/2}}\left(\frac{1}{S_x}+\frac{1}{S_y}\right) - \frac{2}{(t-t')^2} \right],\label{integralaniso_improved_readability}
\end{multline}
where the functions $ S_x(t,t')$ and $S_y(t,t')$ are defined as the solutions of the differential equations with initial conditions
\begin{align}
   \Ddot{S}_x &= - \Omega_x ^2 (t) S_x ,\quad S_x(t,t)=0, \quad\dot{S}_x(t,t)=1, \nonumber \\
   \Ddot{S}_y &= - \Omega_y ^2 (t) S_y ,\quad S_y(t,t)=0, \quad\dot{S}_y(t,t)=1. 
   \label{aniso-differential equations}
\end{align}
By construction, Eq.~\eqref{integralaniso_improved_readability} reproduces Eq.~\eqref{iso_integral} in the limit $\hat{q}_x \to \hat{q}_y$. However, as mentioned above, we were not able to further simplify this integrand in the spirit of \eqref{emission_rate_reformulated} because the integrand in the anisotropic case cannot be easily written as a total derivative.

\subsection{Mean energy loss due to single gluon emission}
Nevertheless, in both cases, we may approximate the mean energy loss of the jet due to a single emission as the mean energy of the emitted gluon,
\footnote{
Performing the integral here 
in the region $\omega\in [0,\infty)$ constitutes another approximation and simplification. As discussed above, our formalism is only valid for $\max(m_s,m_D^4/\qhat)=\omegamin\ll\omega\ll \mathcal E$.
However, since $\mathcal P(\omega)\sim \omega^{-1/2}$ for $\omega\ll \omega_c$, with the characteristic gluon frequency $\omega_c$ from Eq.~\eqref{eq:omegac}, changing the lower integration boundary results in corrections of $\mathcal O(\sqrt{\omegamin/\omega_c})\sim \mathcal O(1/(\alpha_s T L))$. 
Thus, for sufficiently large lengths $L\gg 1/(g^2T)$ these corrections to the typical result $\sim\omega_c$ can be ignored. Note that $1/(g^2T)\sim 1/\Gammael$ is also the typical time between the elastic collisions triggering the splitting process. Since $\mathcal P(\omega)\sim \omega^{-2}$ for $\omega \gg \omega_c$, changing the upper boundary results in an error of $\mathcal O(\omega_c/\mathcal E)\sim \mathcal O(\alpha_s^2T^3 L^2/\mathcal E)$, which can be ignored for sufficiently large parton energies $\mathcal E$.}
\begin{align}
    E_{\text{iso, aniso}}=\int_0 ^\infty \dd \omega \omega \mathcal{P}_{\text{iso, aniso}}(\omega).
    \label{energy-loss-def}
\end{align}
While realistically, jets lose energy via multiple emissions instead of just one, we employ this simplified description to obtain a single number, $E$, allowing us to easily compare and assess the effects of the anisotropy. For a discussion of the relation of this quantity to energy loss due to multiple emissions, we refer to Ref.~\cite{Jeon:2003gi}.
To compare the differences between the isotropic and the anisotropic formalism, we consider the differences in the mean energy loss, which we denote as $\Delta E$,
\begin{equation}
    \Delta E = E_\text{aniso} - E_\text{iso} =\int_0 ^\infty \dd\omega [\omega \mathcal{P}_{\text{aniso}}(\omega)-\omega \mathcal{P}_{\text{iso}}(\omega)].
    \label{Delta-E-def}
\end{equation}
Note that for compactness, we will frequently abbreviate $E\equiv E_\mathrm{iso}$.

\subsection{Special case: A static anisotropic brick of QGP}
We consider the special case of static media of size $L$ characterized by constant $\hat{q}_x$ and $\hat{q}_y$ within the medium, such that \cite{Arnold:2008vd}
\begin{align}
    \omega \mathcal{P}_{\text{iso}}(\omega) &= \frac{2\alpha_s \CR}{\pi}\, \ln|\cos{\Omega L}|\,.
\end{align}
With the characteristic gluon frequency $\omega_c$, 
\begin{align}
\omega_c =\frac{1}{2}\qhat L^2,\label{eq:omegac}
\end{align}
it follows that $\Omega^2 L^2 = -i \frac{\omega_c}{\omega}$, and thus
$E_\mathrm{iso}$ scales linearly with $\omega_c$ \cite{BAIER1998403}
\begin{align}
    E_\text{iso} &= \frac{2\alpha_s \CR}{\pi} \int_0^\infty \dd\omega \ln|\cos{\Omega L}|
    = \chi \alpha_s \CR \omega_c\,,
    \label{isotropic-energy-loss}
\end{align}
with $\chi = 0.5$. 
The same substitutions as for the isotropic system, $\tilde \omega = \omega/\omega_c$, $\tilde t = t/L$, and $\tilde t' = t'/L$, are applicable to the integrals in the anisotropic case for $\omega \mathcal{P}_\text{aniso}(\omega)$ in \eqref{integralaniso_improved_readability} and the differential equations \eqref{aniso-differential equations}. This eliminates any $L$ dependence apart from the same linear scaling $E_\text{aniso} \propto \omega_c$.  
However, in contrast to the isotropic case, $E_\text{aniso}$ additionally depends on the anisotropy $\Delta \hat q/\hat q$, which is a consequence of the substitutions applied to Eqs.~\eqref{aniso-differential equations}.
Moreover, as we have not found an analytic expression for $\omega \mathcal{P}_\text{aniso}(\omega)$ even for a static medium, we will evaluate the integrals numerically. 

Let us now consider Eq.~\eqref{integralaniso_improved_readability} for such an anisotropic static ``brick". To solve the differential equations \eqref{aniso-differential equations}, we have to distinguish between four regions:
\begin{enumerate}
    \item[(a)] $t<L$, \quad $t'<L$\,; \qquad (b) $t<L$, \quad $t'>L$\,;
    \item[(c)] $t>L$, \quad $ t'<L$\,; \qquad (d) $t>L$, \quad $t'>L$\,.
\end{enumerate}
The region (b) does not contribute, since this region is not included in the domain of integration. Additionally, region (d) corresponds to the vacuum region, where we find $S_{x,\,\mathrm{d}}(t,t')=S_{y,\, \mathrm{d}}(t,t')=t-t'$, which cancels in the integrand.
For (a) and (c) we find
\begin{subequations}\label{eq:S_a_functions}
    \begin{align}
        S_{x,\,\mathrm{a}}(t,t') = \Omega_x^{-1} \sin\left(\Omega_x(t-t')\right), \\
        S_{y,\,\mathrm{a}}(t,t') = \Omega_y^{-1} \sin\left(\Omega_y(t-t')\right),
    \end{align}
\end{subequations}
\begin{subequations}
    \begin{align}
         S_{x,\mathrm{c}}(t,t') = \cos\left(\Omega_x(L-t')\right)(t-L) + \Omega_x^{-1} \sin\left(\Omega_x(L-t')\right), \\
        S_{y,\mathrm{c}}(t,t') = \cos\left(\Omega_y(L-t')\right)(t-L) + \Omega_y^{-1} \sin\left(\Omega_y(L-t')\right).
    \end{align}
\end{subequations}
Hence, we obtain the following expression for the anisotropic energy loss of static media with size $L$
\begin{multline}
    E_\text{aniso} = -\frac{\alpha_s \CR}{\pi} \int_0^\infty \dd\omega \; \mathrm{Re} \int_0^L \dd{t'} \\
    \left[ \int_{t'}^L \dd t \left(\frac{1} {S_{x,\mathrm{a}}^{1/2}S_{y,\mathrm{a}}^{1/2}}\left(\frac{1}{S_{x,\mathrm{a}}}+\frac{1}{S_{y,\mathrm{a}}}\right)  - \frac{2}{(t-t')^2} \right) \right. \\
    \left. + \int_L^\infty \dd{t} \left( \frac{1} {S_{x,\mathrm{c}}^{1/2}S_{y,\mathrm{c}}^{1/2}}\left(\frac{1}{S_{x,\mathrm{c}}}+\frac{1}{S_{y,\mathrm{c}}}\right) - \frac{2}{(t-t')^2} \right) \right].
    \label{static-medium-anisotropic-double-integral}
\end{multline}
We evaluate these integrals using extrapolation techniques, where we start from finite $\omega_\mathrm{min}, \, \omega_\mathrm{max}$ and $t_\mathrm{max}$: 

The inner integral of region (c) (bottom line in Eq.~\eqref{static-medium-anisotropic-double-integral})
is evaluated up to a finite $t_\text{max}$, which is then extrapolated numerically for large
$t_\mathrm{max} \to \infty$. Similarly, for the outer integral, we integrate from a carefully%
\footnote{Here, $\omega_\text{min}$ and $\omega_\text{max}$ must be chosen in the regions where the scaling form \eqref{scaling-behaviour-emission-rate} provides a good approximation.} 
chosen $\omega_\text{min}$ up to $\omega_\text{max}$ and then extrapolate using analytic expressions of the spectrum at low and high frequencies given in Appendix \ref{app:asymp_spectrum}. 
Details on our numerical calculations and the extrapolation procedures are provided in Appendix \ref{app:extrapolation}.

\begin{figure}
    \centering
    \includegraphics[width=1\linewidth]{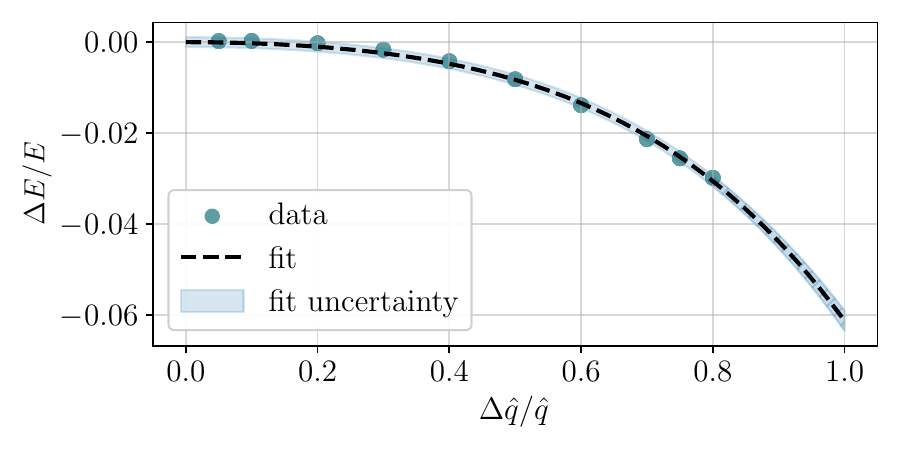}
    \vspace{-1cm}\caption{{Relative difference between isotropic and anisotropic energy loss $\Delta E/E_{\mathrm{iso}} =\frac{E_\text{aniso} - E_\text{iso}}{E_\text{iso}}$ 
    as a function of anisotropy $\Delta\hat{q}/\hat{q}$. The fit using Eq.~\eqref{pocket-formula} is depicted as a dashed line. The anisotropic energy loss $E_\mathrm{aniso}$ is generally smaller than $E_\mathrm{iso}.$ For maximum anisotropy $\Delta\qhat/\qhat = 1$ we find $\Delta E/E \approx -6\%$.}
    } 
    \label{fig:relative-energy-loss-difference}
\end{figure}

\subsection{Numerical results: Mean energy loss in a static anisotropic plasma}
We present our numerical evaluation of Eq.~\eqref{static-medium-anisotropic-double-integral} in
Fig.~\ref{fig:relative-energy-loss-difference}, where we depict the relative energy loss difference $\Delta E/E_\text{iso}$ as a function of anisotropy.  Note again that we approximate the energy loss of the parton here by the mean energy of a single emitted gluon.
As $E_\text{iso}$ and $E_\text{aniso}$ exhibit the same linear scaling with $\omega_c$, as discussed below Eq.~\eqref{isotropic-energy-loss},
their fraction is independent of the medium size $L$.
The relative energy loss difference due to anisotropy, $\Delta E/E_\text{iso}$, 
turns out to be negative and rather small 
(less than $0.2\%$ in magnitude for $\Delta \qhat/\qhat<0.2$). In general, we observe that $E_\text{aniso} < E_\text{iso}$, which is consistent with other recent studies employing the full collision kernel instead of the harmonic approximation \cite{Altenburger:2025iqa, Lindenbauer:2025ctw}.
Even for extremely large anisotropies $\Delta\qhat/\qhat\simeq0.8$, we find that taking into account the anisotropy reduces the mean energy loss by about \qty{3}{\percent}, as compared to using an isotropic $\qhat$.
Evaluating this at a realistic coupling of $\alpha_s=0.3$ and for rather large values of the parameters, i.e., a medium length of $L \simeq \SI{8}{\femto\meter}$, an anisotropy of $\frac{\Delta\hat{q}}{\hat{q}} \simeq 0.75$, and for $\qhat=\SI{5}{\giga\electronvolt^2/\femto\meter}$, we find that the change due to the anisotropy is $\Delta E \simeq \SI{-9}{\giga\electronvolt}$, amounting to only $2\%$ of the total energy loss, making this effect experimentally challenging to measure.
We now further aim to provide a convenient semi-analytic expression for $\Delta E/E$ as a function of the anisotropy $\Delta \qhat/\qhat$.
To this end, we formally expand $E_\text{aniso}$ in $\Delta\hat{q}/\hat{q}$  around an isotropic background $(\hat{q}_x,\hat{q}_y)=((\hat{q}+\Delta\hat{q})/2, (\hat{q}-\Delta \hat{q})/2)$ (see Eq.~\eqref{eq:qhat-comparison-iso}).
Using this symmetric ansatz, any odd powers vanish, such that the energy change due to anisotropy can be expressed as
\begin{align}
    \label{pocket-formula}
   \Delta E &= E_{\text{aniso}} - E_{\text{iso}} \\ 
   &= 
   E_{\text{iso}} 
   \left[a\left(\frac{\Delta\hat{q}}{\hat{q}}\right)^2+b\left(\frac{\Delta\hat{q}}{\hat{q}}\right)^4 +\mathcal{O}\left(\left(\frac{\Delta\hat{q}}{\hat{q}}\right)^6\right)\right]. \nonumber
\end{align}
Neglecting higher-order terms, 
the energy loss for the anisotropic system 
reads
\begin{align} \label{eq:fit_form}
    E_\mathrm{aniso} \approx\chi\alpha_s \CR \omega_c \left[1+a\left(\frac{\Delta\hat{q}}{\hat{q}}\right)^2+b\left(\frac{\Delta\hat{q}}{\hat{q}}\right)^4\right].
\end{align}

We determined the dimensionless parameters $a$ and $b$ 
by fitting Eq.~\eqref{eq:fit_form} to the 
numerically evaluated Eq.~\eqref{static-medium-anisotropic-double-integral}, which resulted in
\begin{align}
    a=-0.0237\,, \qquad b=-0.0376\,,
    \label{eq:fit_values}
\end{align}
with the coefficient of determination $R^2=0.9973$.
Even though we neglect higher-order terms in \eqref{pocket-formula}, this estimate seems to accurately represent the numerical data, as can be seen in Fig.~\ref{fig:relative-energy-loss-difference}. We note again that throughout this paper, our analytical expressions rely on
the $\omega\ll \mathcal E$ limit, such that the expression is independent of the parton energy $\mathcal E$.%
\footnote{The parton energy $\mathcal E$ can, however, have an implicit influence in the applications of this formula via a parton energy dependence of the jet quenching parameter $\qhat$. For instance, this will be the case in the next section where we will use the momentum cutoff model \eqref{eq:cutoff-lpm} that enters logarithmically into $\qhat_i$.}
For maximum anisotropy $\Delta\qhat/\qhat=1$, performing a simple extrapolation using our analytic formula \eqref{pocket-formula} yields a deviation of roughly $6\%$.

Although our derivation and discussions focused on a static medium, we anticipate Eq.~\eqref{pocket-formula} to provide useful results also for time-varying $\qhat$:
Instead of solving Eq.~\eqref{aniso-differential equations} for two different, time-dependent $\qhat_x(t)\neq \qhat_y(t)$, one may decompose the anisotropic mean energy loss into the isotropic part and the difference, $E_\mathrm{aniso} = E_\mathrm{iso}+\Delta E$, and use the \textit{simple formula}
\eqref{emission_rate_reformulated}
for the isotropic part $E_\mathrm{iso}$, while employing the static approximation Eq.~\eqref{pocket-formula} for $\Delta E$.
We expect this approximation to lead to only minor deviations
for a suitable mapping of the time dependent $\qhat$ profile to a corresponding constant $\qhat^{\mathrm{static}}$. It has been shown that a convenient mapping may be given by \cite{Salgado:2002cd}
\begin{align}
    \qhat^{\mathrm{static}}=\frac{2}{L^2}\int_{0}^{L}\dd{t}t\,\qhat(t)\label{eq:qhat-matching}
\end{align}
for an isotropic jet quenching parameter, and we propose to use a similar mapping for the individual components. Such matching procedures to static systems are often considered good approximations and are frequently employed in the literature \cite{Salgado:2002cd, Andres:2016iys, Adhya:2019qse, Andres:2023jao, Mehtar-Tani:2024jtd, Pablos:2025cli}, and we will use them below in our application to expanding nonequilibrium plasmas.

\subsection{Impact on quenching weights\label{sec:quenching-weights}}
\begin{figure}
    \centering
    \includegraphics[width=1\linewidth]{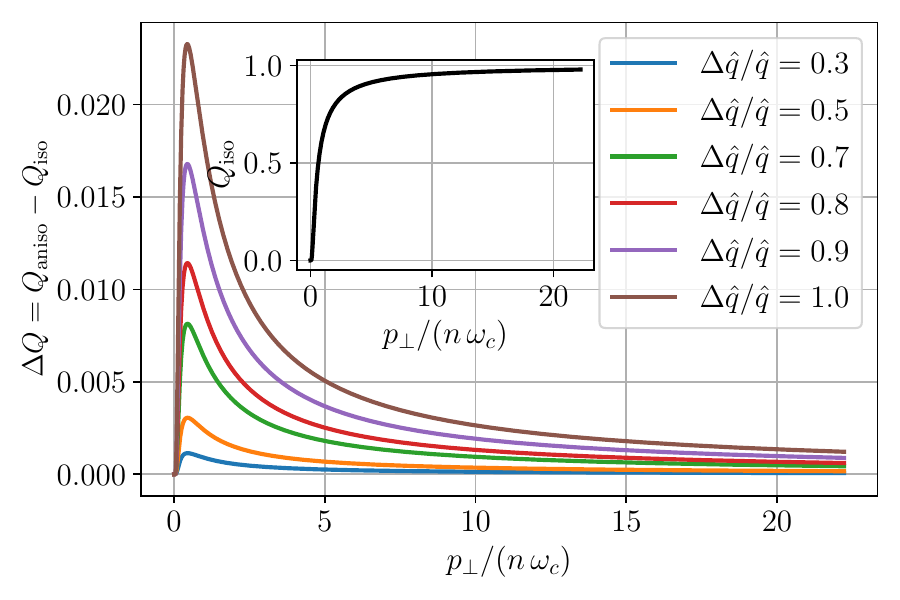}
    \caption{Difference $\Delta Q = Q_\text{aniso}-Q_\text{iso}$ of the quenching weight in an anisotropic ($Q_\text{aniso}$) and corresponding isotropic system ($Q_\text{iso}$) as a function of $p_\perp/(n\omega_c)$. The anisotropy effects decrease with increasing rescaled momentum.
    The inset shows the quenching weight in an isotropic system, which converges to $Q_{\rm iso}\to1$ for large $p_\perp/(n\omega_c)$.
    }
    \label{fig:quenching}
\end{figure}
To calculate the suppression of energetic partons in heavy-ion collisions, it is convenient to introduce the concept of quenching weights \cite{Salgado:2003gb, Baier:2001whe}. They quantify the difference between the parton cross section in vacuum and its medium-modified counterpart,
\begin{align}
    \frac{d \sigma^{{\rm med}}(p_\perp)}{dp_\perp^2} = Q(p_\perp) \, \frac{d \sigma^{{\rm vac}}(p_\perp)}{dp_\perp^2}.
\end{align}
They can be obtained by considering the probability $D(\varepsilon)$ that a parton loses an additional energy $\varepsilon$ when traversing the medium as
\begin{align}\label{eq:Qdef}
Q(p_\perp) = \int \dd\varepsilon\, D(\varepsilon) \left(\frac{\dd{\sigma^{{\rm vac}}}
(p_\perp+\varepsilon)/\dd {p_\perp^2}}{\dd{\sigma^{{\rm vac}}}
(p_\perp)/\dd {p_\perp^2}}\right),
\end{align}
taking into account that a parton with momentum $p_\perp$ was initially produced with momentum $p_\perp + \varepsilon$. Assuming independent emissions, this probability $D(\varepsilon)$ can be expressed as 
\begin{multline}  \label{eq:gluon-prob}
D(\varepsilon) = \sum^\infty_{n=0} \, \frac{1}{n!} \,
\left[ \prod^n_{i=1} \, \int_0^\infty \, \dd{\omega_i} \, \frac{\dd{I(\omega_i)}}{\dd {\omega_i}} 
\right] \\ \delta \left(\varepsilon - \sum_{i=1}^n  \omega_i\right)
\exp \left[ - \int_0^\infty \dd\omega \frac{\dd {I(\omega)}}{\dd\omega} \right]. 
\end{multline} 
Assuming that the particle spectrum falls off with a sufficiently large exponent $n \approx -\frac{\mathrm d}{\dd{ \ln p_\perp}} \ln \frac{\dd{\sigma_\text{vac}}}{\dd{p_\perp^2}}$, the quenching weights can be obtained via
\cite{Baier:2001whe}
\begin{equation}
    \label{eq:Qpperp}
    Q(p_\perp) \approx e^{-nS(p_\perp)/p_\perp}.
\end{equation}
Here, $S(p_\perp)$
is called the shift factor and can be approximated by the mean energy loss $S \approx E$ from Eq.~\eqref{energy-loss-def} for large transverse momenta $p_\perp \gtrsim n\omega_c$.
In analogy to our discussion before, we define the difference between the quenching weights of the anisotropic and the corresponding isotropic medium as $\Delta Q=Q_\text{aniso}-Q_\text{iso}$. It is shown as a function of transverse momentum $p_\perp$ for different jet quenching anisotropies in Fig.~\ref{fig:quenching}, making use of 
the anisotropy expansion \eqref{pocket-formula} with fitted values \eqref{eq:fit_values}.
We find that the anisotropy leads to a small positive contribution to $Q$, i.e., $\Delta Q$ is positive. This is in agreement with decreased energy loss: a larger $Q$ implies less modification of the vacuum cross section. However, the anisotropy only leads to a slight increase of $Q$ of the order of one percent. As expected, particles with larger energy become less modified.

Using $\Delta E\ll p_\perp/n$ leads to
\begin{align}
    \Delta Q/Q_\text{iso} \approx - n\Delta E/p_\perp.
    \label{dQ/Q=dE/p}
\end{align} 
Thus, the relative change to the quenching weight can be estimated from the simple expression \eqref{pocket-formula}.

\section{Rendezvous with $\hat q$ from QCD kinetic theory}\label{sec:kinetictheory}
So far, we have expressed our results for the energy loss and the quenching weights in terms of the plasma anisotropy, measured by the change of the jet quenching parameter in different directions $\Delta \qhat/\qhat$. We will now employ more realistic values for this anisotropy parameter. For that, we use the evolution of the jet quenching parameter $\qhat$ during the early stages in heavy-ion collisions that was numerically computed using QCD kinetic theory in Refs.~\cite{Boguslavski:2023waw, Boguslavski:2024ezg}.
\subsection{Kinetic theory and time scales}
In the QCD kinetic theory description of the plasma background, all information on the plasma time evolution is encoded in the distribution function $f(\vb p,\tau)$, which follows the Boltzmann equation for a homogeneous, longitudinally expanding plasma \cite{Mueller:1999pi, Arnold:2002zm}
\begin{align}
	\pdv[f(\vb p,\tau)]{\tau}- \frac{p_z}{\tau} \pdv[f(\vb p,\tau)]{p_z} =-\Conetwo[f(\vb p,\tau)]- \Ctwotwo[f(\vb p,\tau)],
	\label{eq:boltzmann_equation}
\end{align} 
which can be solved numerically \cite{AbraaoYork:2014hbk, Kurkela:2014tea, Kurkela:2015qoa, kurkela_2023_10409474}. The jet quenching parameter $\qhat_i$ is obtained via \cite{Boguslavski:2023waw}
\begin{align}
	\hat q_{i}=\int_{\substack{q_\perp <\Lambda_\perp\\ p\to \infty}}\dd{\Gamma}\left(q^i\right)^2\left|\mathcal M\right|^2 f(\vb k)\left(1+f(\vb k')\right),
	\label{eq:formula-qhat-explicit}
\end{align}
where medium-effects are included in the gluon-gluon scattering matrix element $\left|\mathcal M\right|^2$ using the isoHTL%
\footnote{For the time evolution, the data provided in \cite{Boguslavski:2024ezg, lindenbauer_2023_10419537} uses Debye-like screening in the matrix elements entering $\Ctwotwo$.} 
screening prescription \cite{Boguslavski:2024kbd}. We use the data provided in \cite{Boguslavski:2024ezg, lindenbauer_2023_10419537} for a Yang-Mills plasma with 't Hooft coupling $\lambda=g^2\NC\in\{0.5,1,2,5,10,20\}$ and perform new simulations for $\lambda\in\{0.25, 0.75, 1.5\}$. For the transverse momentum cutoff $\Lambda_\perp$, we take the LPM cutoff model from Ref.~\cite{Caron-Huot:2008zna, Boguslavski:2024ezg},
\begin{align}\label{eq:cutoff-lpm}
    \Lambda_\perp^{\mathrm{LPM}}(\mathcal E, \Teps)=\zeta^{\mathrm{LPM}} g\times (\mathcal E\Teps^3)^{1/4},
\end{align}
with the jet energy $\mathcal E$ and where the instantaneous temperature $\Teps$ is obtained from the energy density $\varepsilon$ of the kinetic theory simulation via the Landau matching condition
\begin{align}\label{eq:def-Teps}
    \Teps=\left(\frac{30\,\varepsilon}{\nu\pi^2}\right)^{1/4},
\end{align}
i.e., it is the temperature of an equilibrium plasma with the same energy density. The proportionality constant is chosen to be%
\footnote{Ref.~\cite{Boguslavski:2024ezg} includes a typo when referencing the jet energy, which we have corrected here. For $\mathcal E = \qty{100}{\giga\electronvolt}$, one should use $\zeta^{\mathrm{LPM}}=0.70$ for the LPM cutoff.} 
$\zeta^{\mathrm{LPM}}=0.70$.
Here, $\nu=2(\NC^2-1)$ counts the number of gluon degrees of freedom.
Note that in Ref.~\cite{Boguslavski:2024ezg}, the jet direction was chosen to be the $x$-direction, leading to broadening along the $y$ and $z$ directions, which corresponded to $\qhat_{\mathrm{EKT}}^{yy}$ and $\qhat_{\mathrm{EKT}}^{zz}$. In the present study, we align the coordinate system such that the jet moves in the $z$-direction, which implies that we need to identify $\qhat_x=\qhat_{\mathrm{EKT}}^{zz}$ and $\qhat_y=\qhat_{\mathrm{EKT}}^{yy}$.

We initialize the distribution function at time $\tau_0=1/Q_s$ using the initial condition \cite{Kurkela:2015qoa}
\begin{align}
    f(p_\perp,p_z)=\frac{2A\langle p_T\rangle}{\lambda p_\xi}e^{-\frac{2p_\xi^2}{3\langle p_T\rangle^2}},
\end{align}
with $p_\xi=\sqrt{p_\perp^2+(\xi p_z)^2}$ using $\xi=10$, $A=5.24171$, and $\langle p_T\rangle=1.8Q_s$. Here, $Q_s$ is the saturation momentum and should not be confused with the quenching weight $Q$ introduced in Section \ref{sec:quenching-weights}.

It is often convenient to express time in units of the relaxation time 
\begin{align}\label{eq:relaxation-time}
    \tau_R=\frac{4\pi\eta/s}{\Teps},    
\end{align}
which is defined in terms of the specific shear viscosity $\eta/s$ and the effective temperature \eqref{eq:def-Teps}.

Note that while anisotropic QCD plasmas generically develop plasma instabilities \cite{Mrowczynski:1993qm, Mrowczynski:2016etf}, currently they cannot be treated consistently within kinetic theory, and we neglect their effect by employing an isotropic screening prescription \cite{AbraaoYork:2014hbk, Boguslavski:2024kbd}.
For more details on QCD kinetic theory, we refer to \cite{Arnold:2002zm}, and for the implementation we use, to \cite{AbraaoYork:2014hbk, Boguslavski:2023waw, Boguslavski:2024ezg, kurkela_2023_10409474}. The values of the specific shear viscosity used in this paper are given in App.~\ref{app:etas}.

There exists a different timescale associated with the bottom-up thermalization scenario \cite{Baier:2000sb}, where thermalization occurs parametrically at $\tau\sim \tauT$, with
\begin{align}\label{eq:tauT}
    \tauT=\alpha_s^{-13/5}/Q_s.
\end{align}
We will use the symbols $\tilde w$ and $\tilde v$ to denote time rescaled with these timescales, i.e.,
\begin{align}\label{eq:time-rescaling}
    \tilde w=\tau/\tauR, && \tilde v=\tau/\tauT.
\end{align}
The interplay of these time scales in the context of numerical kinetic theory simulations of the bottom-up equilibration process has been discussed in Ref.~\cite{Boguslavski:2023jvg}, where the notion of \emph{limiting attractors} was introduced.
These attractors are obtained by extrapolating observables $\mathfrak O$ to vanishing (weak-coupling bottom-up limiting attractor)
\begin{align}
    \mathfrak O(\tilde v)=\tilde A(\tilde v)+\lambda \, \tilde B(\tilde v)\label{eq:limiting-attractor-general-weak}
\end{align}
and infinite (strong-coupling hydrodynamic limiting attractor) coupling,
\begin{align}
    \mathfrak O(\tilde w)=A(\tilde w)+\frac{1}{\lambda}B(\tilde w),
    \label{eq:strong-limitingattractor-general}
\end{align}
where the extrapolation is performed at fixed rescaled time $\tilde v$ or $\tilde w$. In particular, the weak-coupling limiting attractor was found to provide a good description of the jet quenching parameter ratio $\qhat_x/\qhat_y$ over a large time period and coupling parameter range as well as for different initial conditions \cite{Boguslavski:2023jvg}. 

Note that the approach of different initial conditions to a universal curve for simulations with the same value of the coupling in kinetic theory has been established before \cite{Heller:2016rtz, Almaalol:2020rnu}, which generalizes the earlier notion of a hydrodynamic attractor \cite{Heller:2015dha}. The hydrodynamic limiting attractor emerges when considering how these universal curves for different couplings approach a limiting curve when rescaled with the relaxation time \eqref{eq:relaxation-time}. Similarly, attractors emerging at weak couplings during the last stages in the bottom-up thermalization scenario \cite{Baier:2000sb, Kurkela:2015qoa} lead to a limiting curve of their own, the bottom-up limiting attractor, when rescaled with $\tauT$ \eqref{eq:tauT}.

Before discussing how this applies to energy loss, we reconsider the evolution of the jet quenching parameter in these simulations.

\subsection{Mapping the evolving $\hat q(\tau)$ to a static medium}
While we have described in Sec.~\ref{sec:generalization-anisotropic} how to obtain the spectrum \eqref{static-medium-anisotropic-double-integral} for an arbitrary time evolution of $\hat q_i(\tau)$, for simplicity, our numerical evaluation focused on a static plasma brick with constant $\hat q_i$. For that, we have obtained the simple expression for the difference in the energy loss \eqref{pocket-formula}.
To map the time-evolving system to a static brick, we define the effective jet quenching parameters to be
\begin{subequations}\label{eq:combined-averaging-procedure}
\begin{align}
    \hat{q}_x^\text{eff} (L)&= \frac{2}{L^2} \int_{\tau_0}^{L+\tau_0} \dd\tau(\tau-\tau_0)\,\hat{q}_x(\tau) \,,
    \label{effective-qhatx}\\
    \hat{q}_y^\text{eff} (L) &= \frac{2}{L^2} \int_{\tau_0}^{L+\tau_0} \dd\tau(\tau-\tau_0)\,\hat{q}_y(\tau) \,.
    \label{effective-qhaty}
\end{align}
\end{subequations}
Note that this 
estimate is a minimal anisotropic 
generalization of Eq.~\eqref{eq:qhat-matching}, which is based on the results of Ref.~\cite{Salgado:2002cd} in an expanding plasma with isotropic jet quenching parameter $\qhat_x=\qhat_y$. There, it is shown to provide good agreement for the spectrum and quenching weights. In the diagonal basis of $\qhat_{ij}$, requiring the map to the effective static medium to be linear, to act componentwise, and to reduce to Eq.~\eqref{eq:qhat-matching} for $\qhat_x=\qhat_y$, naturally leads to Eqs.~\eqref{effective-qhatx} and \eqref{effective-qhaty}. Hence, this choice of mapping the individual components not only reduces to the isotropic result \eqref{eq:qhat-matching} for $\qhat_x=\qhat_y$, but additionally preserves the ratio $\qhat_x/\qhat_y$ when the time-dependence can be factorized into a common function, $\qhat_{x,y}(\tau)=f(\tau)a_{x,y}$. In general, the mapping places more weight on later times, diminishing the effect of large anisotropies at early times.
Nevertheless, we emphasize that this procedure \eqref{eq:combined-averaging-procedure} is a modeling choice motivated by the isotropic case. In these expressions, we vary the upper boundary $L$ to model different medium lengths. Effectively, $L$ denotes the maximum distance (time) that the jet travels while quenching occurs. Note that we must be careful when interpreting the results from these effective jet quenching parameters: The resulting curves and spectra do not represent the instantaneous energy loss at length $L$, but rather the mean energy loss of an energetic parton due to a single gluon emission in a medium with an extent $L$.

\begin{figure}
    \centering
    \includegraphics[width=1\linewidth]{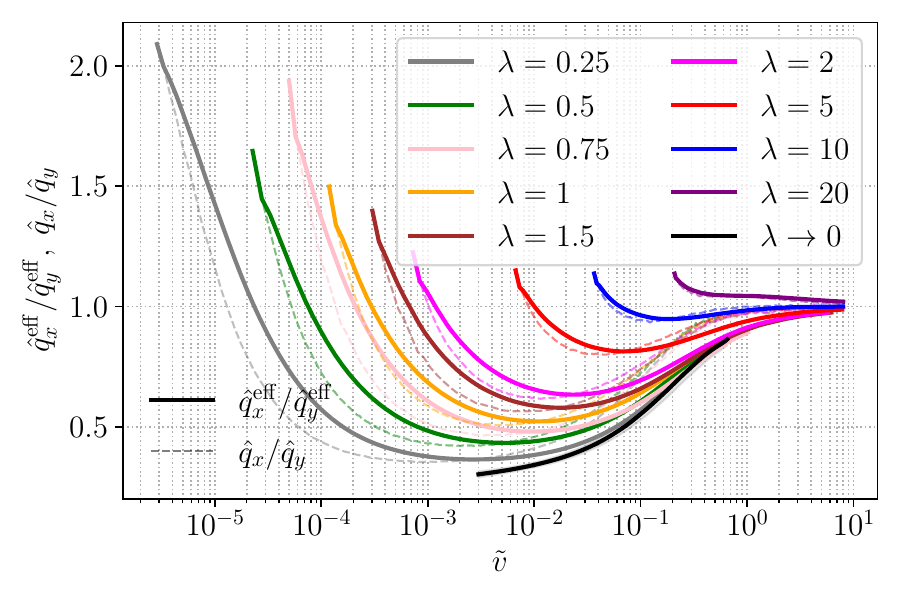}
    \caption{Comparison between the effective jet quenching parameters $\hat{q}_x^\text{eff}/\hat{q}_y^\text{eff}$ (solid lines) as well as the instantaneous jet quenching parameters $\hat{q}_x/\hat{q}_y$ (dashed lines) for various couplings $\lambda$. The curves indicate larger anisotropy at earlier times together with a ``turning point", where the curves pass from $\qhat_x > \qhat_y$ to $\qhat_x < \qhat_y$, and similar for $\qhat_x^\mathrm{eff}/\qhat_y^\mathrm{eff}$.
    The time axis is rescaled with the bottom-up timescale $\tauT$, i.e., $\tilde v=\tau/\tauT$ (Eq.~\eqref{eq:time-rescaling}). We also show the limiting weak-coupling attractor \eqref{eq:limiting-attractor-general-weak} as a black curve denoted by $\lambda \to 0$.
    }
    \label{fig:EKT-effective-qhats}
\end{figure}

\begin{figure}
    \centering
    \includegraphics[width=1\linewidth]{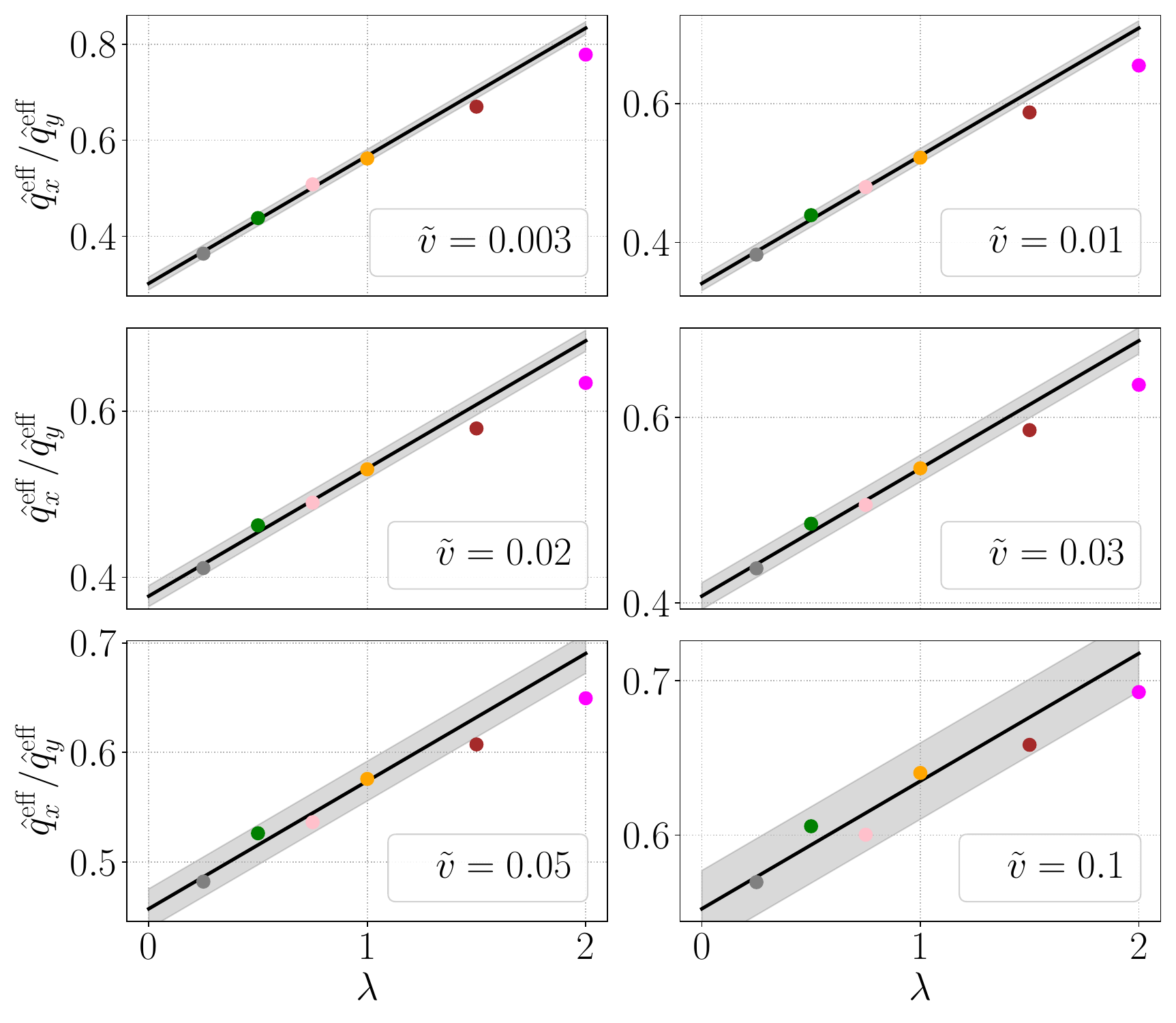}
    \caption{The
    ratio of the effective jet quenching parameters
    $\qhat_x^\mathrm{eff}/\qhat^\mathrm{eff}_y$ for fixed rescaled times $\tilde v=\tau/\tauT$. We also show the linear extrapolation to the weak-coupling limiting attractor \eqref{eq:limiting-attractor-general-weak}.
    }
    \label{fig:fit-for-fixed-times-qhat}
\end{figure}
Figure \ref{fig:EKT-effective-qhats} displays the ratio of the time-averaged jet quenching parameters $\qhat^\mathrm{eff}_x$ and $\qhat^\mathrm{eff}_y$ (solid lines) as a function of $\tilde v=(L+\tau_0)/\tauT$ in comparison to the ratio of the ``instantaneous'' jet quenching parameters $\qhat_x$ and $\qhat_y$ (dashed lines) as a function of $\tilde v=\tau/\tauT$ for different couplings $\lambda$, where $\tau_0=1/Q_s$ is the initial time of the kinetic theory simulation. We observe that the averaged jet quenching parameters exhibit the same qualitative behavior as the instantaneous jet quenching parameters. Additionally, Fig.~\ref{fig:EKT-effective-qhats} reveals an interesting ``crossover point" without anisotropy,  $\qhat_x/\qhat_y=1$ and $\qhat_x^\mathrm{eff}/\qhat_y^\mathrm{eff}=1$ respectively, early in the time evolution. The anisotropy then increases again before the ratio converges to unity in equilibrium. The maximum values of 
$|\Delta \qhat^\text{eff}|/\qhat^\text{eff}$
range from $\approx 0.3$ for $\lambda=0.5$ to below $\lesssim 0.1$ for $\lambda=10$,
where $\Delta \qhat^\text{eff}$ and $\hat q^\text{eff}$ are obtained for every $L$ from Eq.~\eqref{eq:qhat-comparison-iso}. A remarkable feature of the results for the jet quenching parameter $\qhat_i$ from Ref.~\cite{Boguslavski:2024ezg} is the appearance of a weak-coupling \emph{limiting attractor} \cite{Boguslavski:2023jvg}, which emerges when extrapolating the ratio $\qhat_x/\qhat_y$ to vanishing coupling at fixed rescaled time $\tilde v$ using Eq.~\eqref{eq:limiting-attractor-general-weak}.
It can be expected that this limiting attractor related to the anisotropy in $\qhat_x/\qhat_y$ may influence the energy loss ratio, which is obtained from this anisotropy ratio. However, since to obtain the spectrum (and, thus, the energy loss) we need to integrate over the entire time evolution of the jet path, it is a priori unclear how much of this limiting attractor survives the averaging procedure \eqref{eq:combined-averaging-procedure}. First, we consider the effect of the averaging procedure on the limiting attractor.
For that, we additionally show in Fig.~\ref{fig:EKT-effective-qhats} the linear extrapolation of the ratio of the \textit{effective} jet quenching parameters for $\lambda\to 0$,
\begin{align}\label{eq:weakcoupling-qhatratioeff}
    \frac{\qhat_x^\mathrm{eff}}{\qhat_y^\mathrm{eff}}(\tilde v,\,\lambda) = \tilde A_{\qhat}(\tilde v)+\tilde B_{\qhat}(\tilde v)\lambda\,,
\end{align}
obtaining again a \emph{limiting attractor}  $\tilde A_{\qhat}(\tilde v)$ according to Eq.~\eqref{eq:limiting-attractor-general-weak} for sufficiently late times.

The extrapolation procedure is illustrated in Fig.~\ref{fig:fit-for-fixed-times-qhat} for different rescaled times $\tilde v$. For each of these distinct times, we plot the value of $\qhat_x^{\mathrm{eff}}/\qhat_y^{\mathrm{eff}}$ for different couplings, and perform a linear regression to \eqref{eq:limiting-attractor-general-weak}. The error band shows the error estimate of this procedure, i.e., the deviation from the linear form. We observe in Fig.~\ref{fig:fit-for-fixed-times-qhat} that for our couplings, this extrapolation procedure works well for rescaled times $3\times 10^{-3} <\tilde v < 10^{-1}$, thus spanning several orders of magnitude.
We will discuss this attractor in more detail in the next section.

\begin{figure}
    \centering
    \includegraphics[width=1\linewidth]{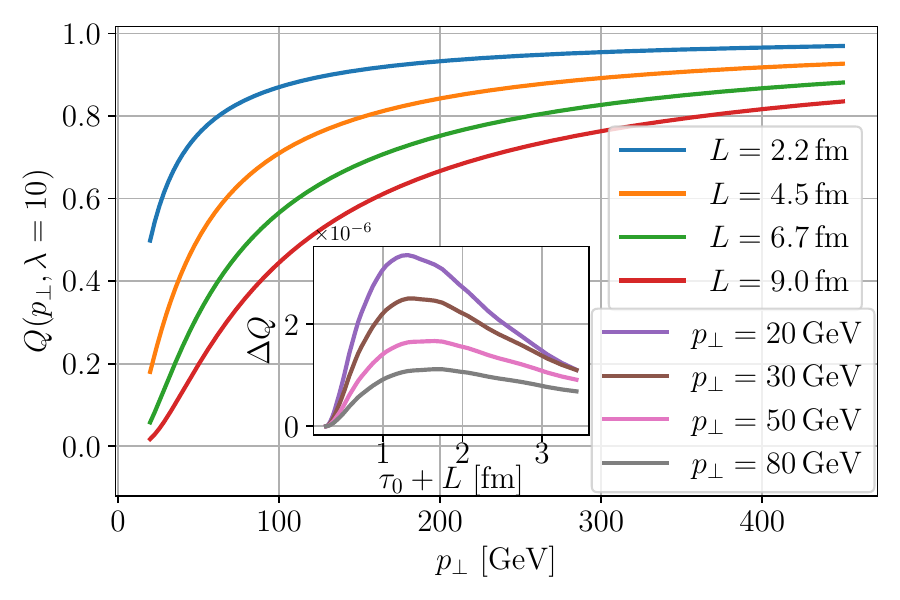}
    \caption{Quenching weight $Q$ as a function of $p_\perp$ for $\lambda=10$ for various fixed medium sizes $L$. For smaller medium sizes, the convergence to $1$ goes more rapidly. The inset shows $\Delta Q=Q_\text{aniso}-Q_\text{iso}$ as a function of medium size $L$ for various transverse momenta $p_\perp$. The impact of the anisotropy mostly increases with larger medium sizes.
    }
    \label{fig:EKT-quenching-weight}
\end{figure}

Next, we discuss the quenching weights introduced in Section \ref{sec:quenching-weights}. We obtain them from the energy loss calculated via \eqref{eq:fit_form} using the averaged jet quenching parameters \eqref{eq:combined-averaging-procedure} as input.
Figure \ref{fig:EKT-quenching-weight} presents the quenching weight $Q$ as a function of the transverse momentum $p_\perp$ for $\lambda=10$, $n=3$, and different medium sizes $L$.
Characteristically, larger medium sizes correspond to smaller quenching weight values, implying larger deviations from the vacuum cross section and, thus, more quenching. Particles with larger transverse momenta $p_\perp$ are less quenched. Figure~\ref{fig:EKT-quenching-weight} also includes the absolute difference $\Delta Q=Q_\mathrm{aniso}-Q_\mathrm{iso}$ as a function of $\tau_0+L$ for $\lambda=10$ and fixed transverse momenta $p_\perp$ in the inset. We obtain only small deviations $\Delta Q\sim 10^{-6}$ for realistic values of $L$.

\begin{figure}
    \centering
    \includegraphics[width=1\linewidth]{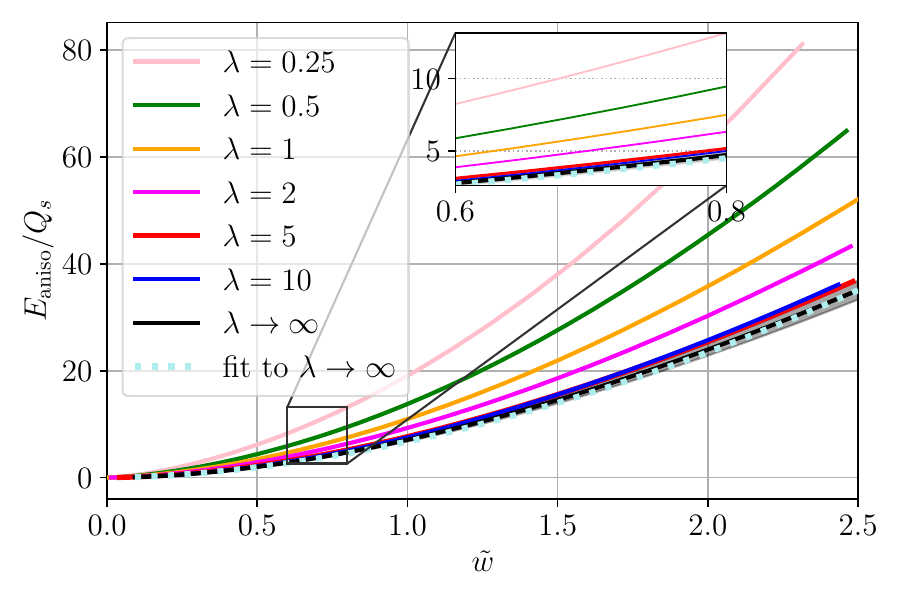}
    \caption{The mean energy loss $E_\text{aniso}$  as a function of the rescaled medium size $\tilde w=(L+\tau_0)/\tauR$.
    We also show the extrapolation $\lambda \to \infty$ to the strong-coupling limiting attractor \eqref{eq:strong-limitingattractor} with an error bar, as well as its fit \eqref{eq:hydro_lim_attractor_fit}.
    Note that equal $\tilde w$ correspond to different medium sizes for different couplings. 
    } 
    \label{fig:KT-E_aniso}
\end{figure}

\subsection{Energy loss meets limiting attractors}

As we have discussed before, the ratio of effective jet quenching parameters ${\qhat_x^\mathrm{eff}}/{\qhat_y^\mathrm{eff}}$ approaches a weak-coupling limiting attractor, which can be obtained by extrapolating  to vanishing coupling at fixed rescaled times $\tilde v$ using Eq.~\eqref{eq:weakcoupling-qhatratioeff}. To study the emergence of limiting attractors in energy loss, we now compare energy loss observables for different couplings $\lambda$ rescaled with appropriate time scales. Recall that we consider the length of the jet path $L+\tau_0$ as the natural time variable, where $\tau_0$ denotes the starting time of quenching, which is the initialization time of the QCD kinetic theory simulations in Ref.~\cite{Boguslavski:2024ezg}.

Figure \ref{fig:KT-E_aniso} displays the mean energy loss $E_\text{aniso}$ as a function of $\tilde w=(L+\tau_0)/\tau_R$ for various couplings. Note that the ordering of the coupling appears to be in a seemingly unintuitive way compared to what one would expect from the scaling law $E\sim L^2$ in a static brick; the smallest coupling shows the largest mean energy loss and the largest coupling the smallest. The reason for this phenomenon is that the medium size also enters in $\tau_R$, which generally grows for decreasing coupling. Hence, the same $(L+\tau_0)/\tau_R$ corresponds to much larger medium sizes for smaller couplings, which explains the observed $L$-dependence.

These curves allow us to extrapolate to large coupling $\lambda\to\infty$ to obtain the hydrodynamic limiting attractor, according to Eq.~\eqref{eq:strong-limitingattractor-general}. More concretely, we take the mean energy loss $E_{\mathrm{aniso}}$ in units of $Q_s$ at a fixed value of the rescaled time $\tilde w=(L+\tau_0)/\tau_R$ as a function of the coupling $\lambda$, and perform a linear regression to
\begin{align}
    \frac{E_{\mathrm{aniso}}}{Q_s}(\tilde w)=A_E(\tilde w)+B_E(\tilde w) \times \frac{1}{\lambda}, \label{eq:strong-limitingattractor}
\end{align}
with the function $A_E(\tilde w)$ identified as the strong-coupling limiting attractor. In practice, we use 
the $\lambda=1, \ 2, \  5, \ 10$ results for the extrapolation, which is shown as a black solid line in Fig.~\ref{fig:KT-E_aniso}. The curves for $\lambda=5$ and $\lambda=10$ almost overlap and nearly coincide with the limiting attractor curve. An error estimate of the regression procedure is also included in the plot. Note that the ordering of the $\lambda=5$ and $\lambda=10$ results seems to be slightly off, which we attribute to late-time discretization effects and numerical uncertainties coming from the extraction of the specific shear viscosity $\eta/s$, as is also discussed in Ref.~\cite{Boguslavski:2024kbd}. Let us now illustrate how this hydrodynamic limiting attractor can be used for simple estimates. We aim to provide a rough estimate of the mean energy loss due to single gluon emission as depicted in Fig. \ref{fig:KT-E_aniso}. For simplicity, we fit the hydrodynamic limiting attractor curve with a power law, 
\begin{align}
    E_{\mathrm{aniso}}^{\mathrm{lim}}/Q_s \approx 6.73\times \tilde w^{1.8},
    \label{eq:hydro_lim_attractor_fit}
\end{align}
with a coefficient of determination $R^2>0.999$.
We will now rewrite this relation in terms of more physical quantities related to different collision systems.
When hydrodynamics becomes applicable, $\langle \tau s\rangle\approx \mathrm{const}$, i.e., no more entropy is produced during a hydrodynamic evolution close enough to ideal hydrodynamics, and the constant will depend on the specific collision system under consideration. Using
Eq.~\eqref{eq:relaxation-time}
and the temperature dependence in ideal hydrodynamics,
\begin{align}
    T=\frac{\Lambda_T}{(\Lambda_T\tau)^{1/3}},
\end{align}
we can derive a simplistic estimate. In a thermal system with $\nu$ degrees of freedom, the entropy density is given by $s=2\nu\pi^2 T^3/45$, which can be used to fix the constant $\Lambda_T^2=\frac{45}{2\pi^2\nu}\langle s \tau\rangle$. The
scale variable is then given by
\begin{align}
    \tilde w=\tau/\tauR=\tau^{2/3}\left(\frac{45}{2\pi^2\nu}\right)^{1/3}\langle s\tau\rangle^{1/3}\frac{1}{4\pi\eta/s}.
\end{align}
Together with the identification $\tau \approx L$, this yields
\begin{align}
    E_\mathrm{aniso}^\mathrm{{lim}}&\approx 6.73\times\left(\frac{45}{2\pi^2}\right)^{3/5}Q_s\nu^{-3/5}L^{6/5}\langle s\tau\rangle^{3/5}\frac{1}{(4\pi\eta/s)^{9/5}}\\
    &\approx \qty{54}{\giga\electronvolt}\times \left(\frac{Q_s}{\qty{1.4}{\giga\electronvolt}}\right)\left(\frac{\langle s \tau\rangle}{\qty{4.1}{\giga\electronvolt^2}}\right)^{3/5}
    \label{eq:estimate}\\
    &\qquad\qquad\times\left(\frac{\nu}{40}\right)^{-3/5}\left(\frac{L}{\qty{5}{\femto\meter}}\right)^{6/5}
    \left(\frac{4\pi\eta/s}{2}\right)^{-9/5}\nonumber
\end{align}
This formula provides a simple way to estimate the mean energy loss of an energetic parton as a function of the medium length $L$, the saturation momentum $Q_s$, the number of degrees of freedom $\nu$, the specific shear viscosity $\eta/s$, and the entropy density (multiplied by time) $\langle \tau s\rangle$, for which we have included typical values, also used in similar estimates in several recent studies
\cite{Keegan:2016cpi, Kurkela:2018vqr, Kurkela:2018wud, Giacalone:2019ldn, Garcia-Montero:2023lrd, Zhou:2024ysb, Garcia-Montero:2024lbl, Boguslavski:2025ylx}.
While Eq.~\eqref{eq:estimate} is a very naive and simplistic estimate, it highlights the simplifying features of attractors and scaling behavior in heavy-ion collisions, and how they may be used in modeling heavy-ion collisions \cite{Kurkela:2018wud, Pablos:2025cli}.

We note that there are several reasons why the resulting estimate in Eq.~\eqref{eq:estimate} might be too large. First, we use the values of $\hat q$ obtained in Ref.~\cite{Boguslavski:2024ezg}, where the proportionality constant in the cutoff in Eq.~\eqref{eq:cutoff-lpm} was fitted to reproduce the value of $\hat q$ from the JETSCAPE collaboration \cite{JETSCAPE:2021ehl}. In this reference, no jet quenching before the hydrodynamic stage was assumed. In contrast, we also consider quenching in the pre-equilibrium stage, and it has been observed previously in the literature that taking this into account requires a smaller value of $\qhat$ when fitting to experimental data
\cite{Andres:2016iys, Andres:2022bql}. Furthermore, we work in the limit in which the energy of the emitted gluon $\omega$ is small with respect to the jet energy, $\omega\ll \mathcal E$. Within this approximation, the parton energy $\mathcal E$ drops out, and equivalently, we consider an infinitely energetic parton. In practice, apart from corrections to the splitting function and to the frequencies $\Omega_i$, the integral over $\omega$ in Eq.~\eqref{static-medium-anisotropic-double-integral} will be bounded by $\mathcal E$, leading to an overall reduction of the resulting energy loss for finite parton energy.

\begin{figure}
    \centering
    \includegraphics[width=1\linewidth]{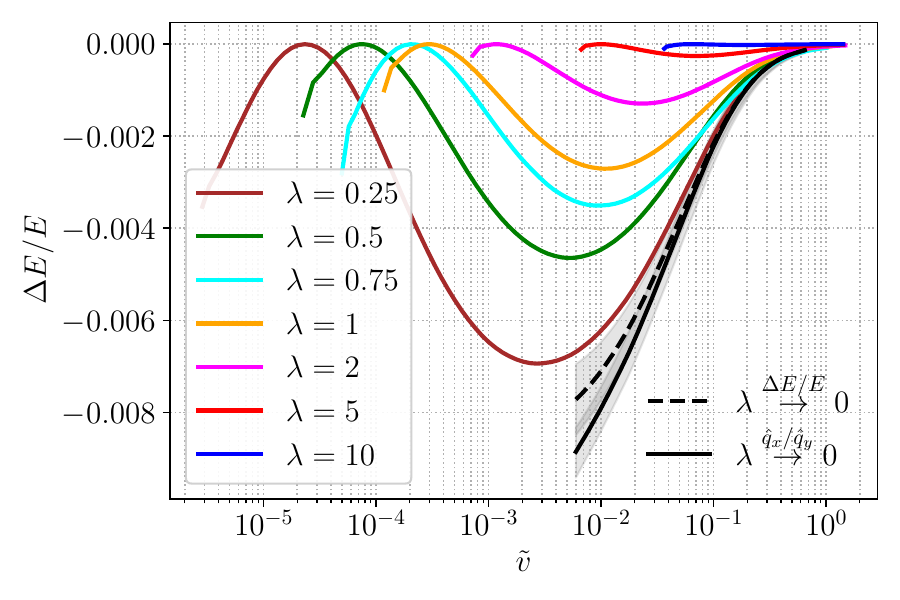}
    \caption{The relative change in energy loss $\Delta E/E$ due to the anisotropy as a function of the medium length $\tilde v =(\tau_0+L)/\tauT$ for different couplings $\lambda$. We also show the extrapolated $\lambda\to 0$ curve, comparing the form from Eq.~\eqref{eq:full-extrapolation-form-for-dE/E}, which we write as $\lambda\overset{\qhat_x/\qhat_y}\to 0$  and an intrinsically linear fit \eqref{eq:limiting-attractor-weak-deltaE-over-E} denoted as $\lambda \overset{\Delta E/E}{\to} 0$.
    }
    \label{fig:lambda-to-0-extrapolation}
\end{figure}

\begin{figure}
    \centering
    \includegraphics[width=1\linewidth]{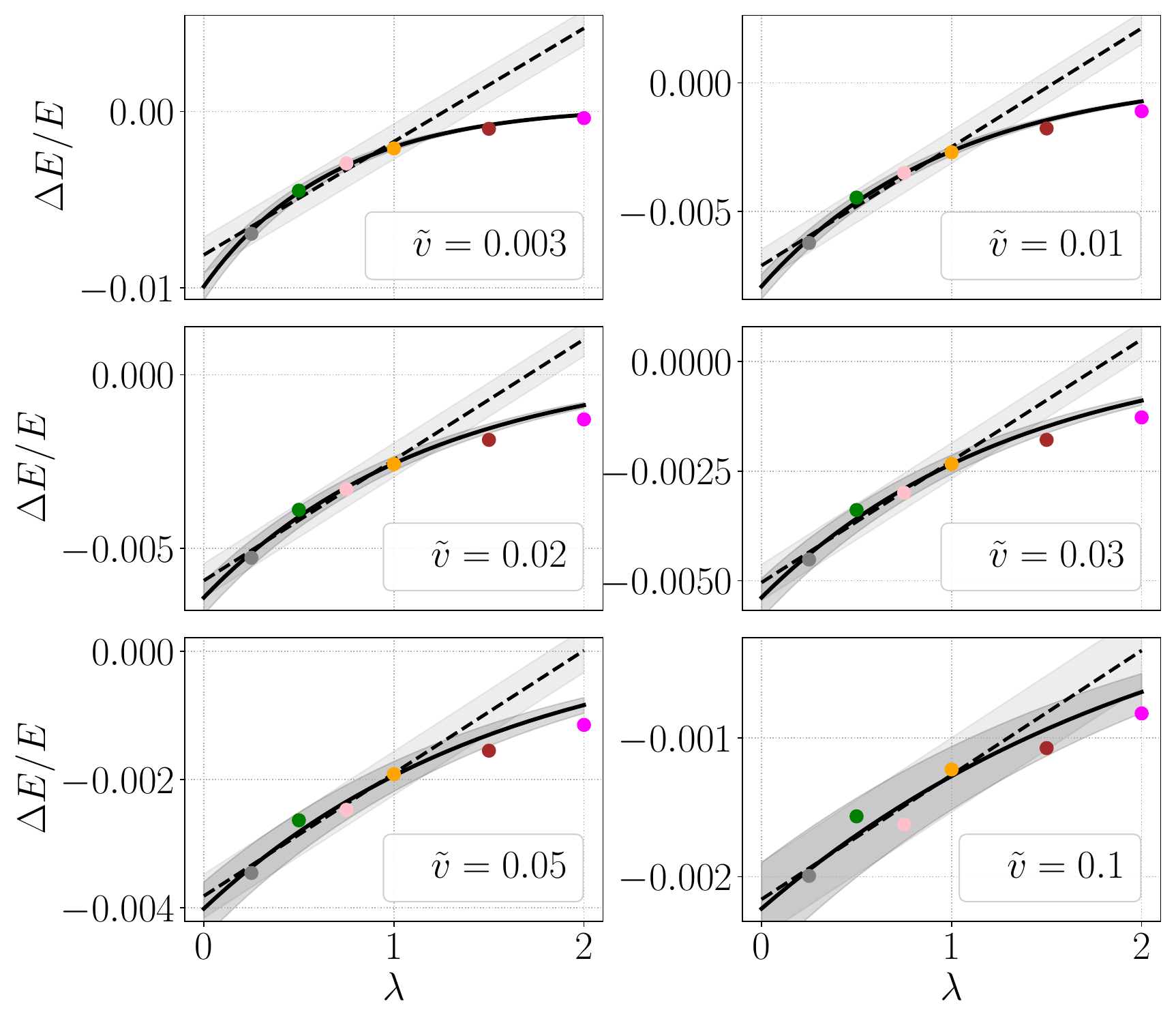}
    \caption{The $\lambda \to 0$ extrapolation for fixed times $\tilde v$ of the relative energy loss $\Delta E/E$ to obtain the bottom-up limiting attractor: We compare a purely linear fit \eqref{eq:limiting-attractor-weak-deltaE-over-E} with the intrinsic fit form \eqref{eq:full-extrapolation-form-for-dE/E}, where we used the limiting attractor \eqref{eq:weakcoupling-qhatratioeff} of the jet quenching parameter ratio $\qhat_x^\mathrm{eff}/\qhat_y^\mathrm{eff}$. 
    }
    \label{fig:fit-for-fixed-times-dE/E}
\end{figure}

We now turn to the relative change in the mean energy loss due to the anisotropy, $\Delta E/E$, which we show in Fig.~\ref{fig:lambda-to-0-extrapolation} as a function of the rescaled time $\tilde v=(L+\tau_0)/\tauT$.
We perform the extrapolation to the weak-coupling bottom-up limiting attractor, using
\begin{align}
    \frac{\Delta E}{E}(\tilde v)=\tilde A_{\Delta E}(\tilde v)+ \tilde B_{\Delta E}(\tilde v)\times \lambda,\label{eq:limiting-attractor-weak-deltaE-over-E}
\end{align}
and identifying $\tilde A_{\Delta E}(\tilde v)$ as the weak-coupling bottom-up limiting attractor.
Although the fitting procedure using the couplings $\lambda\in\{0.25,\,0.5,\, 0.75,\, 1\}$ results in a larger uncertainty than for the limiting attractor in the $\hat q^{\mathrm{eff}}$ ratio of Fig.~\ref{fig:EKT-effective-qhats}, the extrapolation leads to a well-defined curve, which is shown as a black dashed line in the bottom panel. The large error bar of the extrapolation reflects the fact that the relative energy loss difference is badly described within a linear relation to the ratio of the effective jet quenching parameters. A more precise estimate for the weak-coupling limiting attractor of $\Delta E/E$ can be obtained from Eq.~\eqref{pocket-formula} with the limiting attractor found for the jet quenching parameter ratio $\hat q_x^{\mathrm{eff}}/\hat q_y^{\mathrm{eff}}$ in \eqref{eq:weakcoupling-qhatratioeff},
\begin{align}
    {\Delta E}/{E}(\lambda) = a \left(\frac{\tilde A_{\qhat}+\tilde B_{\qhat}\lambda -1}{\tilde A_{\qhat}+\tilde B_{\qhat}\lambda+1}\right)^2 + b\left(\frac{\tilde A_{\qhat}+\tilde B_{\qhat}\lambda -1}{\tilde A_{\qhat}+\tilde B_{\qhat}\lambda+1}\right)^4.
\label{eq:full-extrapolation-form-for-dE/E}
\end{align}
The resulting limiting attractor,
\begin{align}
    \tilde {\mathcal A}=a\left(\frac{\tilde A_{\qhat}-1}{\tilde A_{\qhat}+1}\right)^ 2+b\left(\frac{\tilde A_{\qhat}-1}{\tilde A_{\qhat}+1}\right)^4
\end{align}
is shown as a black solid line in Fig.~\ref{fig:lambda-to-0-extrapolation}.
While at very small coupling one retrieves a linear dependence, the nonlinear terms quickly become important. Both of these extrapolation procedures are illustrated in Fig.~\ref{fig:fit-for-fixed-times-dE/E}. There, we show the relative mean energy change due to the anisotropy $\Delta E/E$ as a function of the coupling $\lambda$ for various distinct rescaled times $\tilde v$. We show both the results and error estimates for the linear fit \eqref{eq:limiting-attractor-weak-deltaE-over-E} (dashed line) and the more complete nonlinear result \eqref{eq:full-extrapolation-form-for-dE/E}. The linear fit slightly overestimates the value at $\lambda \to 0$. However, we observe that both approaches lead to similar results with overlapping uncertainties. We have therefore shown that the recent concept of \emph{limiting attractors} appears in simple jet energy loss calculations and thus drawn it closer to phenomenological applications.
\section{Conclusion}\label{sec:conclusion}
In this work, we have studied how jets lose energy when traversing an anisotropic plasma.
Using the harmonic approximation,
we have computed the corrections to the mean energy of an emitted gluon due to the anisotropy, which is an important ingredient for jet energy loss calculations. We found that when comparing with the corresponding isotropic system, the anisotropy in the jet quenching parameter $\Delta \hat q/\hat q$ leads to relative changes in the mean energy loss $\Delta E/E$ due to a single gluon emission of less than six percent. By fitting the resulting values, we have obtained a simple expression for this quantity given by Eq.~\eqref{pocket-formula}.

Using the time evolution of the jet quenching parameter recently extracted during the initial stages in heavy-ion collisions, we estimated the resulting mean energy loss during the early stages within this formalism.
In particular, we first mapped the time-dependent and anisotropic jet quenching parameters $\qhat_x(\tau)\neq\qhat_y(\tau)$ onto a static anisotropic plasma.
We then showed that both the anisotropic energy loss $E_\mathrm{aniso}$ and, while still small in size, the relative change in energy loss due to the anisotropy $\Delta E/E$ follow the recently introduced limiting attractors, obtained by extrapolation to infinite and vanishing coupling, respectively. This universal behavior allowed us to propose a simple estimate of the mean energy loss of a jet given by Eq.~\eqref{eq:estimate}.
Our findings imply that the anisotropy during the early stages of a heavy-ion collision leaves only a modest imprint on the mean energy of emitted gluons. 
Therefore, in order to test anisotropy effects, more differential observables that, for example, depend on the azimuthal angle should be considered. A natural next step would be to consider the entire time dependence of the jet quenching parameter and go beyond the static brick limit employed in the present study, which we leave for future work. In particular, it will be interesting to examine whether this influences the emergence of limiting attractors. This will also be interesting in the context of more differential observables. Such universal behavior may provide useful input to phenomenological models or Monte Carlo event generators.
\section*{acknowledgments} 
We thank A.~Altenburger, C.~Andres, J.~Barata, A.~Kudinor, A.~Kurkela, T.~Lappi, J.~Peuron, K.~Rajagopal, and A.~Sadofyev for engaging discussions, useful comments, and collaboration on related projects.
This work is funded in part by the Austrian Science Fund (FWF) under Grant DOI 10.55776/P34455 and 10.55776/J4902 (FL). FL is a recipient of a DOC Fellowship of the Austrian Academy of Sciences at TU Wien (project 27203).
    For the purpose of open access, the authors have applied a CC BY public copyright license to any Author Accepted Manuscript (AAM) version arising from this submission.
The results in this paper have been achieved in part using the Austrian Scientific Computing (ASC) infrastructure, project 71444.

\section*{Data availability}
Our implementation and data are publicly available \cite{horl_2026_18931819}.

\appendix

\section{DGLAP Splitting Functions\label{sec:dglap}}
The DGLAP splitting functions, which enter Eq.~\eqref{emissionrategeneral}, are given by \cite{Arnold2:2008}
\begin{align}
    P_{\text{q} \rightarrow \text{g}}(x) &= \CF \frac{1+(1-x)^2}{x},\\
    P_{\text{g} \rightarrow \text{g}}(x) &= \CA \frac{1+x^4+(1-x)^4}{x(1-x)},
\end{align}
where $x$ is the energy fraction of the emitted gluon. For $x\ll 1$, they can be simplified to $P_{\text{s} \rightarrow \text{g}}(x) \approx 2\CR/x$, with $\CR=\CA$ or $\CR=\CF$ depending on the particle species $s$.

\section{Asymptotic behavior of the spectrum}
\label{app:asymp_spectrum}
In a static brick, the gluon emission probability is given by
\begin{align}
    \omega \mathcal{P}(\omega) = \frac{2\alpha_sC_\mathrm{R}}{\pi}\ln \left|\cos (\Omega(\omega) L)\right|,
\end{align}
where $\Omega^2L^2=-i\frac{\omega_c}{\omega}$. It is straightforward to consider the limiting cases of small and large gluon energies $\omega$,
\begin{subequations}\label{scaling-behaviour-emission-rate}
\begin{align}\label{eq:omega-small}
    \omega \mathcal{P}_\text{iso}(\omega) \propto \omega ^{-1/2} \quad \text{for} \quad \omega \ll \omega_\mathrm{c}\,, 
\end{align}
\begin{align}\label{eq:omega-large}
    \omega \mathcal{P}_\text{iso}(\omega) \propto \omega^{-2} \quad \text{for} \quad \omega \gg \omega_\mathrm{c}\,.
\end{align}
\end{subequations}
For $|\Omega| L\ll 1$, we can use a simple Taylor expansion to recover \eqref{eq:omega-large}. In the opposite case $|\Omega| L\gg 1$, we can write $\cos(\Omega L) \sim e^{i\Omega L}$, which results in \eqref{eq:omega-small} after application of the logarithm.

In the anisotropic case, we do not have such a compact formula, but we numerically find good agreement with these analytic limits.
These scaling laws are useful for our numerical implementation, as outlined in the following section.
\section{Numerical Implementation of the spectrum}
\label{app:extrapolation}
\begin{figure}[t]
    \centering
    \includegraphics[width=1\linewidth]{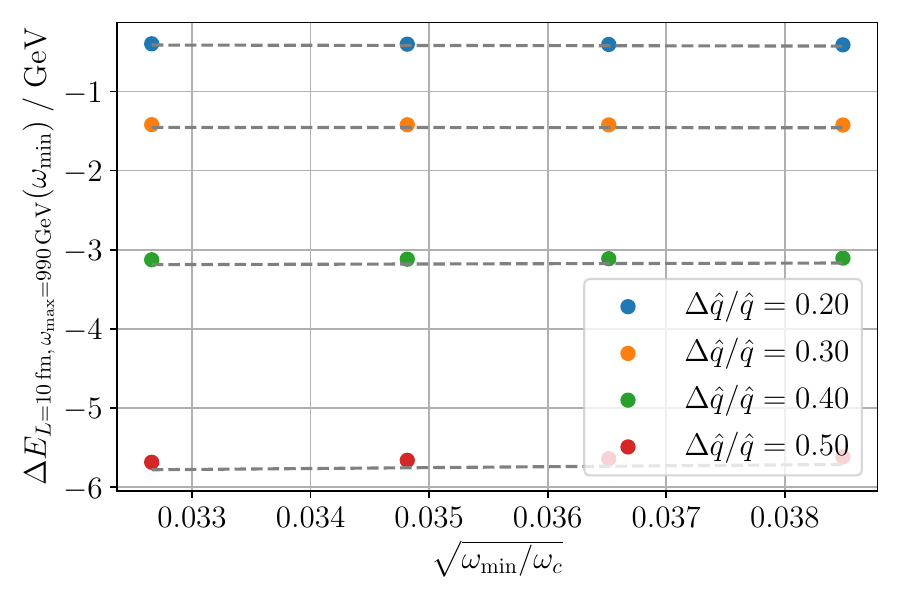}
    \includegraphics[width=1\linewidth]{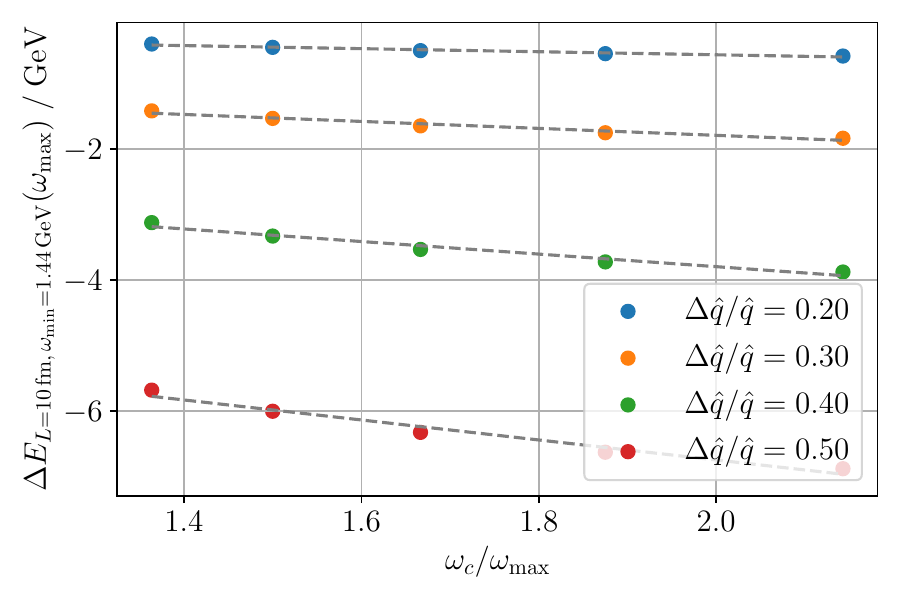}
    \caption{Two projections of the extrapolation procedure for the boundaries $\omega_\mathrm{min}$ and $\omega_\mathrm{max}$ using Eq.~\eqref{eq:DE_wmin_wmax}.
    {\em (Top:)} Extrapolation for fixed $\omega_\mathrm{max}$ and $L$ as a function of $\omega_\mathrm{min}$. $\Delta E$ is observed to be roughly independent of the lower integral bound $\omega_\mathrm{min}$.
    {\em (Bottom:)} Extrapolation for fixed $\omega_\mathrm{min}$ and $L$ as a function of $\omega_\mathrm{max}$. $\Delta E$ depends as expected on the inverse power of $\omega_\mathrm{max}$.
    }
    \label{fig:wmin-wmax-extrapolation}
\end{figure}

\begin{figure}[t]
    \centering
    \includegraphics[width=1\linewidth]{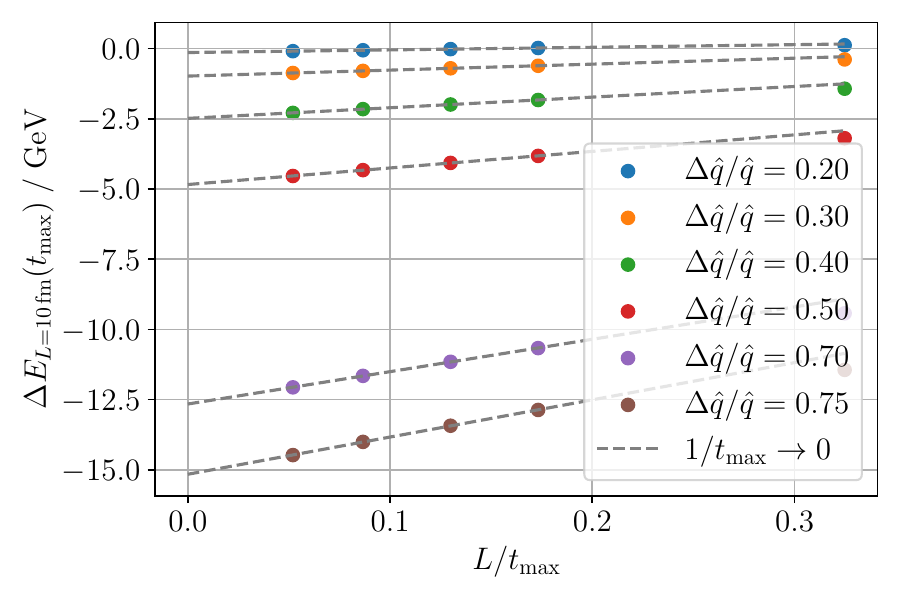}
    \caption{Evaluation of the inner integral \eqref{approximate-inner-integral} with $L=10 \ \text{fm}$ and $\hat{q}=1 \ \text{GeV}^3$ for $t_\text{max}=160, \ 300, \ 400, \ 600, \ 1000 \ \text{GeV}^{-1}$ and the corresponding extrapolation $\frac{1}{t_\text{max}}\to 0$.
    }
    \label{fig:tmax-extrapolation}
\end{figure}
\begin{figure}[t]
    \centering
    \includegraphics[width=1\linewidth]{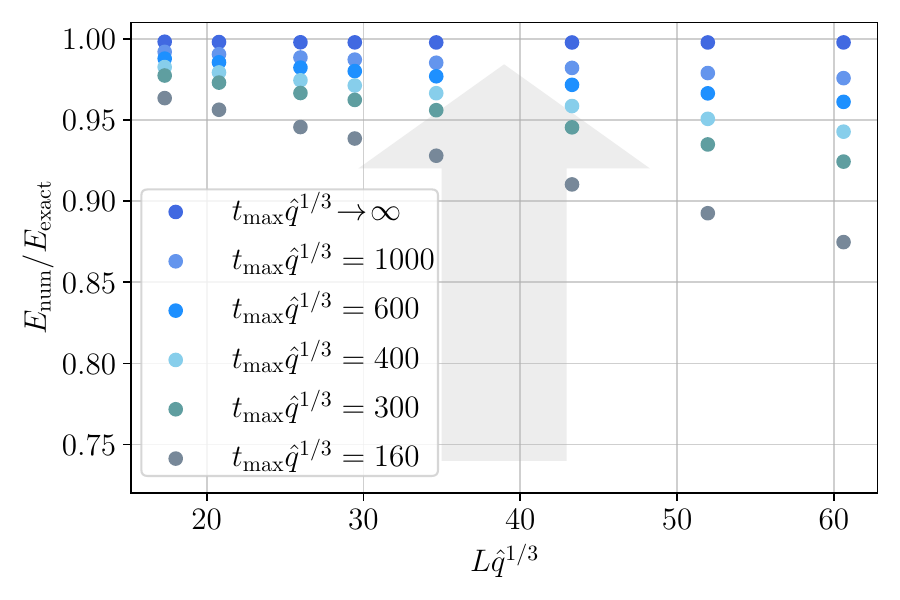}
    \caption{The energy loss $E_\text{iso}(L)$ in an isotropic system calculated from \eqref{static-medium-anisotropic-double-integral}, and compared to \eqref{isotropic-energy-loss} for different $t_\text{max}$ and  $\hat{q}=1 \ \text{GeV}^3$. Using a linear extrapolation (see Fig.~\ref{fig:tmax-extrapolation}), we recover the analytic result \eqref{isotropic-energy-loss}.
    }
    \label{fig:compare-isotropic-energy-loss}
\end{figure}
Here, we describe how we numerically obtain the spectrum in practice. Our implementation is openly accessible \cite{horl_2026_18931819}.
We directly compute Eq.~\eqref{static-medium-anisotropic-double-integral} using a standard numerical integrator.
The inner integral
\begin{align}
\int_{L}^{\infty} \dd t \ \left[ \frac{1}{S_{x,\mathrm{c}}^{1/2}S_{y,\mathrm{c}}^{1/2}}\left(\frac{1}{S_{x,\mathrm{c}}}+\frac{1}{S_{y,\mathrm{c}}}\right) - \frac{2}{(t-t')^2} \right]
\end{align}
requires special treatment due to the upper integration boundary. We regularize it, evaluating the finite integral
\begin{align}
    \int_{L}^{t_\text{max}} \dd t \ \left[ \frac{1}{S_{x,\mathrm{c}}^{1/2}S_{y,\mathrm{c}}^{1/2}}\left(\frac{1}{S_{x,\mathrm{c}}}+\frac{1}{S_{y,\mathrm{c}}}\right) - \frac{2}{(t-t')^2} \right]
    \label{approximate-inner-integral}
\end{align}
for $t_\text{max}$ mainly ranging from $300\ \mathrm{GeV}^{-1}$ to $1000 \ \mathrm{GeV}^{-1}$, where a linear extrapolation in $\frac{1}{t_\text{max}}$ is possible. Moreover, we approximate the outer integral \eqref{Delta-E-def} as
\begin{equation}
    \Delta E = E_\text{aniso} - E_\text{iso} =\int_{\omega_\text{min}} ^{\omega_\text{max}} \dd\omega [\omega \mathcal{P}_{\text{aniso}}(\omega)-\omega \mathcal{P}_{\text{iso}}(\omega)].
\end{equation}
We then perform a two dimensional extrapolation of the form, 
\begin{align}
    \label{eq:DE_wmin_wmax}
    \Delta E(\omega_\text{min}, \omega_\text{max}) = \Delta E+A\omega_\text{min}^{+1/2}+B\omega_\text{max}^{-1},
\end{align}
stemming from the integration of the assumed scaling laws \eqref{scaling-behaviour-emission-rate}. In particular, we fit $\Delta E$, $A$ and $B$ simultaneously for each anisotropy $\Delta \hat q/\hat q$ and fixed $t_\text{max}$, which corresponds to sending $\omega_\text{min} \to 0$ and $\omega_\text{max} \to \infty$. 
This is illustrated in Fig.~\ref{fig:wmin-wmax-extrapolation}, where we show the $\omega_\text{min}$ and $\omega_\text{max}$ extrapolations in the top and bottom panel, respectively. Note that, while the fitting procedure seems to work well, the $\omega_\mathrm{min}$ and $\omega_\mathrm{max}$ extrapolations induce a small systematic shift of the data points for large anisotropies visible in the panels.
The $t_\mathrm{max}$ extrapolation is conducted subsequently and shown in Fig.~\ref{fig:tmax-extrapolation}. 
To verify the quality of our numerical results, and particularly the $t_\mathrm{max}$ and previous extrapolations, we computed the isotropic energy loss using Eq.~\eqref{static-medium-anisotropic-double-integral} for finite $t_\text{max}$ and compared it to the analytical form \eqref{isotropic-energy-loss}. As depicted in Fig.~\ref{fig:compare-isotropic-energy-loss}, we accurately retrieve the analytical exact results for this case.

\begin{table}[th]
\begin{ruledtabular}
  \centering
  \label{tab:lambda-eta-s}
  \begin{tabular}{c c c c c c c c c c }
    $\lambda$ & 0.25 & 0.5 & 0.75 & 1 & 1.5 & 2 & 5 & 10 & 20 \\
    \hline
    $\eta/s$ & 280 & 76.5 & 39.5 & 22.6 & 12 & 7.08 & 1.58 & 0.55 & 0.20 \\
  \end{tabular}
  \end{ruledtabular}
    \caption{Numerical values of the specific shear-viscosity $\eta/s$ used in this paper for different couplings $\lambda$.}
    \label{tab:etas}
\end{table}

\section{Values of the specific shear viscosity}\label{app:etas}

To obtain the relaxation time \eqref{eq:relaxation-time}, we need the value of the specific shear viscosity $\eta/s$, which, in general, is a function of the coupling $\lambda$. For several couplings, this value was extracted and tabulated in Refs.~\cite{Keegan:2015avk, Boguslavski:2024kbd}. Since we perform simulations with additional couplings, we follow the same procedure as in Ref.~\cite{Boguslavski:2024kbd} to extract the corresponding values of the specific shear viscosity $\eta/s$. 
In particular, it can be obtained from the late-time behavior of the pressure anisotropy,
\begin{align}
    \frac{P_L}{P_T}=1-\frac{2\tauR}{\tau\pi},
\end{align}
where the pressure ratio is obtained from
\begin{align}
    P_L=\nu\int\frac{\dd[3]{\vb p}}{(2\pi)^3}\frac{p_z^2}{p}f(\vb p), && P_T=\nu\int\frac{\dd[3]{\vb p}}{(2\pi)^3}\frac{p_x^2}{p}f(\vb p).
\end{align}
The resulting numerical values used in the present work are provided in Table~\ref{tab:etas}.

\bibliography{bib}
\end{document}